\documentstyle[12pt]{article}

\def\ATMP#1#2#3{{\it Adv. Theor. Math. Phys.} {\bf #1} {(#2)} {#3}}
\def\JHEP#1#2#3{{\it JHEP} {\bf #1} {(#2)} {#3}}
\def\NPB#1#2#3{{\it Nucl. Phys.} {\bf B#1} {(#2)} {#3}}
\def\PLB#1#2#3{{\it Phys. Lett.} {\bf B#1} {(#2)} {#3}}
\def\CQG#1#2#3{{\it Class. Quantum Grav.} {\bf #1} {(#2)} {#3}}
\def\MPLA#1#2#3{{\it Mod. Phys. Lett.} {\bf A#1} {(#2)} {#3}}
\def\RMP#1#2#3{{\it Rev. Mod. Phys.} {\bf #1} {(#2)} {#3}}
\def\PRD#1#2#3{{\it Phys. Rev.} {\bf D#1} {(#2)} {#3}}
\def\PR#1#2#3{{\it Phys. Rep.} {\bf #1} {(#2)} {#3}}
\def\AP#1#2#3{{\it Ann. Phys.} {\bf #1} {(#2)} {#3}}

\renewcommand{\thefootnote}{\fnsymbol{footnote}}
\newcommand{\newsection}{\setcounter{equation}{0}\section}
\def\appendix#1{\addtocounter{section}{1}
\setcounter{equation}{0}\renewcommand{\thesection}{\Alph{section}}
\section*{Appendix \thesection\protect\indent \parbox[t]{11.15cm}{#1}}
\addcontentsline{toc}{section}{Appendix \thesection\ \ \ #1}}

\begin{document}
{}~\hfill FIAN/TD/19-99

{}~\hfill hep-th/9906217

\vspace{3cm}

\begin{center}
{\Large\bf Light-cone form of field dynamics
in anti-de Sitter space-time and AdS/CFT correspondence}

\vspace{2cm}
R.R. Metsaev\footnote{E-mail: metsaev@td.lpi.ac.ru}

\vspace{1cm}
{\it Department of Theoretical Physics, P.N. Lebedev Physical
Institute, Leninsky prospect 53, 117924, Moscow, Russia}

\vspace{3cm}
{\bf Abstract}
\end{center}

\bigskip

Light-cone form of field dynamics in anti-de Sitter space-time
is developed. Using field theoretic and group theoretic approaches the
light-cone representation for generators of anti-de Sitter algebra acting
as differential operators on bulk fields is found.  We also present
light-cone reformulation of the boundary conformal field theory
representations. Making use of these explicit representations of AdS
algebra as isometry algebra in the bulk and the algebra of conformal
transformations at the boundary a precise correspondence
between the bulk fields and the boundary operators is established.

\bigskip
\bigskip
\bigskip


\bigskip
PACS-99:  11.30-j; 11.25.Hf

\bigskip

Published in: Nucl.Phys. B563 (1999) 295-348

\renewcommand{\thefootnote}{\arabic{footnote}}
\setcounter{footnote}{0}

\newpage

\newsection{Introduction}

In spite of its Lorentz noncovariance, the light-cone formalism \cite{DIR}
offers conceptual and technical simplifications of approaches  to  various
problems of modern quantum field theory.
For example,  one can mention the construction of light-cone string field
theory \cite{KK}-\cite{schwarzvol} and superfield formulation for some
versions of supersymmetric theories [6--9].
Sometimes,  a theory formulated within
this formalism turns out to be a good starting point for deriving a
Lorentz covariant formulation \cite{SIEG,HATA}. Another
attractive application of the light-cone formalism is a construction of
interaction vertices in the theory of higher spin massless fields
\cite{BEN1}-\cite{M2}. Some interesting applications of light-cone
formalism to field theory like QCD are reviewed in \cite{BROD}.
Discussion of super $p$-branes and string bit models
in the light-cone gauge is given in \cite{dewit,bergshoeff5} and
\cite{thorn} respectively.

Until now the light-cone formalism was explored in
Minkowski space-time (for a review see \cite{SIEG}).
The  major goal of this paper is to develop
light-cone form of field dynamics in anti-de Sitter space time.
A long term motivation comes from a number of the following potentially
important applications.

One important application is to type IIB
superstring in $AdS_5\times S^5$ background. Motivated by conjectured
duality between the string theory and ${\cal N}=4$, $d=4$ SYM theory
\cite{mal}-\cite{w} the Green-Schwarz formulation of this  string theory
was suggested in \cite{mt1} (for  further developments
see \cite{krr}-\cite{kt}).
Despite considerable efforts these strings have not yet been quantized
(some related
interesting discussions are in
\cite{Dolan:1999pi}-\cite{Berenstein:1999jq}).
As is well known,  quantization of GS superstrings propagating
in flat space is straightforward only in the light-cone gauge.
It is the light-cone gauge that removes unphysical
degrees of freedom  explicitly
and reduces the action to quadratical form in
string coordinates.
The light-cone gauge in string theory implies the corresponding
light-cone formulation for target space fields.
In the case of strings in AdS background this suggests
that we should first study a light-cone form dynamics of {\it target space
fields}
propagating in AdS space-time. Understanding  a light-cone
description of AdS target space fields might help to solve problems of
strings in AdS space-time.

The second application is to a  theory of higher massless
spin fields propagating in AdS space-time. Some time ago completely
self-consistent interacting equations of motion for higher
massless fields of all spins in four-dimensional AdS space-time
have been discovered \cite{vas1}. For generalization to higher
space-time dimensions see \cite{vas2}. Despite efforts the action
that leads to these equations of motion has not yet been obtained.
In order to quantize these theories
and investigate their ultraviolet behavior it would be important
find an appropriate action. Since the higher massless spin
theories correspond
 quantum mechanically
to
non-local point particles
in a  space of certain auxiliary variables, it is conjectured  that
they may be ultraviolet finite (see \cite{vas3},\cite{vas4}).
We believe that a light-cone
formulation is what is required to understand these theories better.
The situation here may be analogous to that in string theory;
a covariant formulation of closed string field theories
is non-polynomial and is not useful for practical calculations,
while  the light-cone  formulation restricts the
string action to cubic order in string fields.

Keeping in mind these extremely important applications, in this paper we
develop and apply the light-cone formalism to the study of AdS/CFT
correspondence at the level state/operators matching.  As is well  known
in  the case of the massless fields, investigation of AdS/CFT
correspondence requires analysis of some subtleties related to the fact
that transformations of bulk massless fields are defined up to local gauge
transformations. These complications are absent in the light-cone
formulation because here we only deal with physical fields and this
allows us to demonstrate AdS/CFT correspondence in a rather
straightforward way.

In this paper we will focus on the light-cone formalism for
integer  (bosonic) arbitrary spin massless fields that can be formulated in
any dimension. There is a number of reasons for this.  First, the massless
case is the simplest one  to demonstrate all essential new features of
light-cone formalism in AdS space-time.  Second, higher spin massless field
theory  is an interesting and important subject which  itself  should be
studied in detail. We develop also light-cone formulation for massive
fields but in  this case we do not discuss the  AdS/CFT correspondence.
Since the method we use is algebraic in nature,  an extension of our
results to the case of half integer spin fields (fermions) is
straightforward and will be done elsewhere.   A  generalization to the
case supersymmetric theories is relatively straightforward and will be
studied in future.

The paper is organized as follows. In section 2 we describe various forms
of the $so(d-1,2)$ algebra and explain our notation.

In section 3 we develop light-cone formulation by using field theoretic
approach. First  we discuss the  simplest case of spin one Maxwell field.
Next we generalize our discussion to the case of totally symmetric and
antisymmetric fields.  We find an explicit representation for generators of
anti-de Sitter algebra acting as differential operators on fields
of arbitrary spin.

Based on these results in section 4 we establish defining equations for
the  AdS algebra generators  for    arbitrary symmetry type fields. We
 demonstrate that  a field in $d$ dimensional AdS space-time can
be considered as a  massive particle in $(d-1)$ flat space time with
a continuous mass spectrum.  We explicitly match the AdS algebra generators
acting on bulk fields with conformal algebra generators acting on fields
with continuous mass spectrum.  This interrelation is behind AdS/CFT
correspondence but it also suggests an idea of how massless higher spin
fields in AdS space time and string theory at the boundary could be
related. We shall briefly comment on  this point.

In section 5 we use group theoretic approach to solve the defining
equations. In this section we develop light-cone form of AdS generators
 for both massless and massive fields. We establish close
correspondence between
field theoretic and group theoretic approaches.

For comparison of boundary values of bulk light-cone fields with operators
of boundary conformal theory we need to develop a light-cone formulation of
conformal theories too. To our knowledge this formulation was not
previously given in the  literature. In section 6 we present  light-cone
formulation of conformal field theory for the case of arbitrary spin
totally symmetric operators. We  consider  operators with
canonical dimension as well as their conformal partners (which
are  sometimes referred to
 as shadow operators or sources).
 We investigate correspondence between solutions of equations
of motion for totally symmetric bulk fields and operators of boundary
conformal theory.  We demonstrate that normalizable modes of bulk field
are related to conformal operators while non-normalizable modes are
related to conformal partners of these conformal operator.

Section 7 summarizes our conclusions and suggests directions for future
research. Appendices contain some mathematical details
and useful formulae.

\newsection{Various forms of $so(d-1,2)$ algebra and notation}

First let us discuss the forms of AdS algebra, that is $so(d-1,2)$,
we are going to use.  AdS algebra of $d$ dimensional AdS
space-time consists of
translation generators $\hat{P}^A$ and rotation
generators $\hat{J}^{AB}$ which span $so(d-1,1)$ Lorentz algebra. The
commutation relations of AdS algebra are

$$
[\hat{P}^A,\hat{P}^B]=\lambda^2\hat{J}^{AB}\,,
\qquad
[\hat{J}^{AB},\hat{J}^{CE}]=\eta^{BC}\hat{J}^{AE}
+3\  \hbox{terms}\,,
$$
$$
[\hat{P}^A,\hat{J}^{BC}]=\eta^{AB}\hat{P}^C-\eta^{AC}\hat{P}^B\,,
$$
$$
\eta^{AB}=(-,+\ldots,+)\,,
\qquad
A,B,C,E=0,1,\ldots,d-1\,.
$$
The $\lambda$ is a cosmological constant of AdS
space-time. Throughout this paper we use antihermitean form of
generators: $G^\dagger=-G$. As $\lambda\rightarrow 0$
the AdS algebra becomes the Poincar\'e algebra

$$
\lim_{\lambda\rightarrow 0} \hat{P}^A=P_{Poin}^A\,,
\qquad
\lim_{\lambda\rightarrow 0} \hat{J}^{AB}=J_{Poin}^{AB}\,.
$$
This form algebra is not convenient for our purposes.
We prefer to use the form provided by nomenclature
of conformal algebra. Namely we introduce new basis

\begin{equation}
\begin{array}{ll}
P^a\equiv \hat{P}^a+\lambda\hat{J}^{d-2a}
&
\hbox{ new translation generators}\,,
\\[9pt]
K^a\equiv \frac{1}{2}(-\frac{1}{\lambda^2} \hat{P}^a
+\frac{1}{\lambda}\hat{J}^{d-2 a})
\qquad &
\hbox{ conformal boost generators}\,,
\\[9pt]
D\equiv -\frac{1}{\lambda}\hat{P}^{d-2}
&
\hbox{ dilatation generator}\,,
\\[9pt]
J^{ab}\equiv \hat{J}^{ab}
&\hbox{ generators  of } so(d-2,1) \hbox{ algebra}\,.
\end{array}
\end{equation}
Flat space limit in this notation is given by

\begin{equation}\label{contract2}
\lim_{\lambda\rightarrow 0}
P^a=P_{Poin}^a\,,
\qquad
\lim_{\lambda\rightarrow 0}
(-\lambda D)=P_{Poin}^{d-2}\,,
\qquad
\lim_{\lambda\rightarrow 0}
(\frac{1}{2\lambda}P^a +\lambda K^a)
=J_{Poin}^{d-2 a}\,.
\end{equation}
In the conformal algebra basis one has the following
well known commutation relations

\begin{eqnarray}
\label{ppkk}
&
{}[D,P^a]=-P^a\,,
\qquad
[D,K^a]=K^a\,,
\qquad
[P^a,P^b]=0\,,
\qquad
[K^a,K^b]=0\,,
&
\\
&
{}[P^a,J^{bc}]=\eta^{ab}P^c
-\eta^{ac}P^b\,,
\qquad
[K^a,J^{bc}]=\eta^{ab}K^c-\eta^{ac}K^b\,,
\\
\label{pkjj}
&
{}[P^a,K^b]=\eta^{ab}D-J^{ab}\,,
\qquad
[J^{ab},J^{ce}]=\eta^{bc}J^{ae}+3\hbox{ terms}\,,
&
\\
&
\eta^{ab}=(-,+,\ldots+)\,,
\qquad
a,b,c,e=0,1,\ldots,d-3,d-1\,.
&
\end{eqnarray}
In this form the AdS algebra is known as the algebra of
conformal transformations in $(d-1)$-dimensional Minkowski
space-time. In sections 3-5 we shall be interested in realization of
this algebra as the one of transformations of massless bulk
fields propagating in $d$-dimensional AdS space-time while in section 6 we
shall realize this algebra as the algebra of conformal transformations on
appropriate operators.

Throughout this paper we shall use Poincar\'e parametrization of AdS
space-time in which

$$
ds^2=\frac{1}{z^2}(-dt^2+dx_i^2+dz^2+ dx_{d-1}^2)\,,
\qquad\qquad
z>0\,.
$$
Here and below we set cosmological constant $\lambda$ equal to unity.
The boundary at spatial infinity corresponds to $z=0$.
\footnote{Poincar\'e coordinates cover half of AdS space-time. Because we
are interested in infinitesimal transformation laws of physical fields
the global description of AdS space-time is not important for our study.}
The Killing vectors in these coordinates are given by\footnote{The
target space indices $\mu,\nu$ take the values $0,1,\ldots,d-1$.}

\begin{eqnarray}\label{kilvec}
\xi^{P^a,\mu}=\eta^{a\mu}\,,
\quad
\xi^{K^a,\mu}=-\frac{1}{2}x_\nu^2\eta^{a\mu}+x^a x^\mu\,,
\quad
\xi^{D,\mu}=x^\mu\,,
\quad
\xi^{J^{ab},\mu}
=x^a\eta^{b\mu} - x^b\eta^{a\mu}\,,
\end{eqnarray}
while the corresponding generators are defined as
$G=\xi^{G,\mu}\partial_\mu$. To develop light-cone formulation we
introduce light-cone variables $x^\pm\,, x^I$ where we use
the convention

$$
x^\pm \equiv\frac{1}{\sqrt{2}}(x^{d-1}\pm x^0)\,,
\qquad
x^I=x^i,\,z;
\qquad
x^0\equiv t,\,\,\,x^{d-2} \equiv z\,,
$$

$$
I,J,K,L=1,\ldots,d-2\,,
\qquad
i,j,k,l=1,\ldots,d-3\,.
$$
In this notation scalar product of tangent space vectors is  decomposed as

$$
X^A Y^A = X^+Y^- + X^-Y^+ +X^IY^I\,,
\qquad
X^IY^I=X^iY^i+X^zY^z\,,
$$
i.e. we use the convention $X^{d-2}=X^z$. The coordinate $x^+$ is
considered as an evolution parameter. Here and below to simplify our
expressions we will drop the metric tensors $\eta_{AB}$, $\eta_{ab}$ in
scalar products.

Because we are going to describe the fields in the
light-cone gauge let us discuss light-cone form of the above algebra.
In light-cone formalism the AdS algebra splits into generators

\begin{equation}\label{kingen}
P^+,\,P^i,\, J^{+i},\, K^+,\, K^i,\, D,\, J^{+-},\,J^{ij}\,,
\end{equation}
which we refer to as kinematical generators and

\begin{equation}\label{dyngen}
P^-,\, J^{-i}\,, K^-\,,
\end{equation}
which we refer to as dynamical generators. For $x^+=0$ the kinematical
generators are realised quadratically in physical fields while the
dynamical generators receive corrections in interaction theory. In this
paper we deal with free fields.  The light-cone form of AdS algebra can be
obtained from (\ref{ppkk})-(\ref{pkjj}) with the light-cone metric having
the following non
vanishing elements $\eta^{+-}=\eta^{-+}=1$, $\eta^{ij}=\delta^{ij}$.

Instead of target space tensor fields we prefer to use tangent space
tensor fields. To pass to tangent space tensor fields we should introduce
local frame. One convenient choice is specified by the frame
one-forms $e^A=e^A_\mu dx^\mu$ with

$$
e_\mu^A=\frac{1}{z}\delta^A_\mu\,.
$$
The connection one-forms, defined by
$de^A+\omega^{AB}\wedge e^B=0$, are then given by

$$
\omega^{AB}_\mu=\frac{1}{z}(\delta^A_z\delta^B_\mu
-\delta^B_z\delta^A_\mu)\,.
$$
Tangent space tensor fields are defined in terms of the  target
space  ones as

\begin{equation}\label{tantar}
\Phi^{A_1\ldots A_s}
\equiv e_{\mu_1}^{A_1}\ldots e_{\mu_s}^{A_s}
A^{\mu_1\ldots \mu_s}\,.
\end{equation}
To simplify our expressions we introduce creation and annihilation
operators $\alpha^A$, $\bar{\alpha}^A$ and construct Fock space vector

\begin{equation}
\label{genfun}
|\Phi\rangle
=\Phi^{A_1\ldots A_s}\alpha^{A_1}\ldots\alpha^{A_s}|0\rangle\,,
\qquad
\bar{\alpha}^A|0\rangle=0\,.
\end{equation}
To describe totally antisymmetric fields we use anticommuting
oscillators\footnote{Throughout this paper we use a convention
$\{x,y\}=xy+yx$.}

\begin{equation}\label{antosc}
\{\bar{\alpha}^A,\alpha^B\}=\eta^{AB}\,,
\qquad
\{\alpha^A,\alpha^B\}=0\,,
\qquad
\{\bar{\alpha}^A,\bar{\alpha}^B\}=0\,,
\end{equation}
while for description of totally symmetric fields we use commuting
oscillators

\begin{equation}\label{comosc}
[\bar{\alpha}^A,\alpha^B]=\eta^{AB}\,,
\qquad
[\alpha^A,\alpha^B]=0\,,
\qquad
[\bar{\alpha}^A,\bar{\alpha}^B]=0\,.
\end{equation}
The Lorentz covariant derivative for $|\Phi\rangle$ takes the form

\begin{equation}\label{lorspiope}
D_\mu\equiv
\partial_\mu+\frac{1}{2}\omega_\mu^{AB}M^{AB}\,,
\qquad
M^{AB}=\alpha^A\bar{\alpha}^B-\alpha^B\bar{\alpha}^A\,.
\end{equation}
where $M^{AB}$ is spin operator of Lorentz algebra $so(d-1,1)$.
In the sequel we often use  the notation

\begin{equation}\label{conv}
\alpha D\equiv \alpha^A D^A\,,
\qquad
\alpha \hat{\partial}\equiv \alpha^A \hat{\partial}^A\,,
\qquad
D_A\equiv e_A^\mu D_\mu\,,
\qquad
\hat{\partial}_A\equiv e^\mu_A\partial_\mu\,,
\qquad
\hat{\partial}^2\equiv\hat{\partial}_A\hat{\partial}_A\,.
\end{equation}
Also,  we adopt  the following conventions for derivatives
$\partial^+=\partial_-$, $\partial^-=\partial_+$,
$\partial^I=\partial_I$, where
$\partial_\pm\equiv \partial/\partial x^\pm$,
$\partial_I\equiv \partial/\partial x^I$.

\newsection{Light-cone formulation of field dynamics in
AdS space-time. Field theoretical approach}

In this section we shall develop light-cone formulation of field dynamics
in AdS space-time by applying field theoretic approach.  The basic strategy,
which is well known, consists of the following steps.  First we start with
gauge invariant equations of motion for free fields in AdS background. We
shall impose light-cone gauge, solve the constraints, and derive equations
of motion for physical degrees of freedom. Next we shall use the original
global AdS group transformations of gauge fields supplemented by
compensating gauge transformation to maintain the gauge. From these we
shall get realization of AdS algebra on the space of physical degrees
of freedom.

\subsection{Maxwell field. Light-cone form of equations of motion}

As a warm up let us consider spin one Maxwell field. Instead of $A^\mu$
with the equations of motion $D_\mu F^{\mu\nu}=0$ we introduce tangent
space field $\Phi^A$ defined by (\ref{tantar}) and use the following form
for equations of motion in tangent space

\begin{equation}\label{eqmot1}
D_BF^{BA}=0\,,
\qquad
F_{AB}=D_A \Phi_B - D_B\Phi_A\,,
\end{equation}
where $F^{AB}$ is the field strength in the tangent space while
$D_A$ is covariant derivative

$$
D_A\Phi_B=\hat{\partial}_A\Phi_B+\omega_{ABC}\Phi_C\,.
$$
In Poincar\'e coordinates one has

$$
D_A\Phi_B=\hat{\partial}_A\Phi_B
+\delta_B^z\Phi_A-\eta_{AB}\Phi_z
$$
and field strength takes the form
$$
F_{AB}=\hat{\partial}_A \Phi_B-\hat{\partial}_B\Phi_A
+\delta_B^z\Phi_A-\delta_A^z\Phi_B\,.
$$
The equations (\ref{eqmot1}) can be cast into the form

$$
\hat{\partial}_BF^{BA}+(2-d)F^{zA}=0
$$
and one has then the following second order equations of motion for the
gauge field $\Phi^A$

\begin{equation}\label{eqmot3}
(\hat{\partial}^2+(1-d)\hat{\partial}_z+d-2)\Phi^A
-\hat{\partial}^A(\hat{\partial}\Phi)
+(d-3)\hat{\partial}^A \Phi^z
+(2-d)\delta_z^A \Phi^z+2\delta_z^A(\hat{\partial}\Phi)=0\,,
\end{equation}
where $\hat{\partial}\Phi\equiv \hat{\partial}^A\Phi^A$.
Since these equations are invariant with respect to the gauge
transformation
$\delta \Phi^A =\hat{\partial}^A\Lambda$
we can impose the light-cone gauge

\begin{equation}\label{lcv}
\Phi^+=0\,.
\end{equation}
Inserting this into equations (\ref{eqmot3}) we get the following
constraint\footnote{Recall that in the Minkowski space-time the Maxwell
equations in gauge $\Phi^+=0$ lead to the Lorentz constraint $\partial^A
\Phi^A=0$.  This is not the case in AdS space-time. Here, by virtue of the
relation $D_A\Phi^A=\hat{\partial}\Phi+(1-d)\Phi^z$, the constraint
(\ref{gcon}) does not coincide with the Lorentz constraint $D^A\Phi^A=0$.}

\begin{equation}\label{gcon}
\hat{\partial}^A\Phi^A=(d-3)\Phi^z\,.
\end{equation} From (\ref{gcon}) we express the $\Phi^-$
in terms of the physical field\footnote{Here as well as while obtaining
the constraint (\ref{gcon}) we assume, as usual in light-cone formalism,
that the  operator $\partial^+$ has trivial kernel.}

\begin{equation}\label{Phmi}
\Phi^- = -\frac{\partial^I}{\partial^+}\Phi^I
+\frac{d-3}{\hat{\partial}^+}\Phi^z\,.
\end{equation}
Note that the second term in r.h.s. of equation (\ref{Phmi}) is absent
in flat space.  It is this term that breaks $so(d-2)$ manifest invariance
and reduce it to $so(d-3)$ one. By virtue of the constraint
(\ref{gcon}) the equations of motion (\ref{eqmot3}) take the form

$$
(\hat{\partial}^2+(1-d)\hat{\partial}_z+d-2)\Phi^A
+(d-4)\delta_z^A \Phi^z=0\,.
$$ From this we get the following equations for the physical fields
$\Phi^i$, $\Phi^z$:

$$
(\hat{\partial}^2+(1-d)\hat{\partial}_z+d-2)\Phi^i=0\,,
\qquad
(\hat{\partial}^2+(1-d)\hat{\partial}_z+2d-6)\Phi^z=0\,.
$$
Since this form of equations of motion is not convenient we introduce
new physical field $\phi^I$ defined by\footnote{Note that it is the field
$\phi^I$ that has conventional canonical dimension $\Delta_0=(d-2)/2$.}

\begin{equation}\label{Phphv}
\Phi^I=z^{(d-2)/2}\phi^I\,.
\end{equation}
In terms of $\phi^I$ the equations of motion take the form

\begin{eqnarray}
\label{eqmotphii}
&&
(\partial^2-\frac{1}{4z^2}(d-2)(d-4))\phi^i=0\,,
\\
\label{eqmotphiz}
&&
(\partial^2-\frac{1}{4z^2}(d-4)(d-6))\phi^z=0\,.
\end{eqnarray}
Dividing by $\partial^+$ these equations can be rewritten in the
Schr\"odinger form

$$
\partial^-\phi^I =P^-\phi^I\,,
$$
where the action of $P^-$ on physical fields is defined by

\begin{eqnarray}
\label{hamvec1}
&&
P^-\phi^i=\Bigl(-\frac{\partial_I^2}{2\partial^+}
+\frac{(d-2)(d-4)}{8z^2\partial^+}\Bigr)\phi^i\,,
\\
\label{hamvec2}
&&
P^-\phi^z=\Bigl(-\frac{\partial_I^2}{2\partial^+}
+\frac{(d-4)(d-6)}{8z^2\partial^+}\Bigr)\phi^z\,.
\end{eqnarray} From equations of motion
(\ref{eqmotphii}), (\ref{eqmotphiz}) we see
that in $d=4$ the mass like terms cancel.  This fact reflects the conformal
invariance of spin one field in four dimensional AdS space-time.
Gauge invariant action for spin one Maxwell field

$$
S=-\frac{1}{4}\int d^dx\sqrt{g}F_{AB}^2
$$
takes the following form in terms of physical field $\phi^I$

\begin{equation}\label{lcac}
S_{l.c.}=\int d^dx \partial^+\phi^I(-\partial^-+P^-)\phi^I\,.
\end{equation}

\subsection{Light-cone form of transformations for spin one physical
degrees of freedom}

Now let us consider transformation laws of physical field
$\phi^I$. Toward this end we start, as usual, with original global  AdS
symmetries, supplemented by compensating gauge transformation required
to maintain the gauge

\begin{equation}\label{tot0}
\delta_{tot}A^\mu
={\cal L}_\xi A^\mu +\partial^\mu \Lambda\,,
\end{equation}
where $\xi^\mu$ are the AdS target space Killing vectors and
${\cal L}$ is the Lie derivative given by

$$
{\cal L}_\xi A^\mu
=\xi^\nu\partial_\nu A^\mu
-A^\nu\partial_\nu \xi^\mu\,.
$$
In terms of tangent space field $\Phi^A$ (\ref{tantar})
and tangent space Killing vectors\footnote{We use the notation $\xi^+$,
$\xi^-$, $\xi^I$ and $\eta^+$, $\eta^-$, $\eta^I$ to indicate the Killing
vectors in target space and tangent space respectively.}

\begin{equation}\label{etxi}
\eta^A\equiv e_\mu^A\xi^\mu
\end{equation}
the transformations (\ref{tot0}) take the form

\begin{equation}\label{tot}
\delta_{tot}\Phi^A={\cal L}_\eta\Phi^A+\hat{\partial}^A\Lambda\,,
\end{equation}
where ${\cal L}_\eta$ is given by

$$
{\cal L}_\eta \Phi^A =\eta^BD_B \Phi^A-\Phi^B D_B \eta^A\,.
$$
In Poincar\'e coordinates we use one has the representation

$$
{\cal L}_\eta \Phi^A
=(\eta\hat{\partial})\Phi^A
+\frac{1}{2}(\hat{\partial}^A\eta^B
-\hat{\partial}^B\eta^A)\Phi^B
+\frac{1}{2}\delta_z^A(\Phi\eta)-\frac{1}{2}\eta^A\Phi^z\,,
$$
where $\eta\hat{\partial}\equiv \eta^A\hat{\partial}^A$,
$\eta\Phi\equiv \eta^A\Phi^A$.
As usual, the gauge parameter $\Lambda$ can be found from the
requirement that the complete transformations (\ref{tot})
maintain the gauge (\ref{lcv}), i.e. from the following equation

$$
\delta_{tot}\Phi^+=0\,.
$$
Solution to the equation for $\Lambda$ is found to be

$$
\Lambda=-\frac{\partial^+\eta^I}{\partial^+}\Phi^I\,.
$$
Now plugging this $\Lambda$ and $\Phi^-$ given in (\ref{Phmi}) into
$\delta_{tot}\Phi^I$ we find transformation laws for the
physical field:

\begin{eqnarray*}
\delta_{tot}\Phi^I
&=&(\eta\hat{\partial})\Phi^I
+\frac{1}{2}(\hat{\partial}^I\eta^J-\hat{\partial}^J\eta^I)\Phi^J
+\frac{1}{2}\delta_z^I\eta^J\Phi^J-\frac{1}{2}\eta^I\Phi^z
\\
&+&\frac{1}{\partial^+}(\hat{\partial}^+\eta^I\partial^J
-\hat{\partial}^+\eta^J\partial^I)\Phi^J
+\frac{d-2}{2\partial^+}(\delta_z^I\partial^+\eta^J
-\delta_z^J\partial^+\eta^I)\Phi^J
\\
&-&\frac{d-4}{2\partial^+}
(\delta_z^I\partial^+\eta^J+\delta_z^J\partial^+\eta^I)\Phi^J
-\frac{1}{\partial^+}\partial^+\eta^z \Phi^I\,,
\end{eqnarray*}
or in terms of $\phi^I$ (\ref{Phphv}) and $\xi^\mu$ (\ref{etxi})

\begin{eqnarray*}
\delta_{tot}\phi^I
&=&(\xi\partial)\phi^I+\frac{d-2}{2z}\xi^z
+\frac{1}{2}(\partial^I\xi^J-\partial^J\xi^I)\phi^J
+\frac{1}{\partial^+}(\partial^+\xi^I\partial^J
-\partial^+\xi^J\partial^I)\phi^J
\\
&-&\frac{d-4}{2z\partial^+}
(\delta_z^I\partial^+\xi^J+\delta_z^J\partial^+\xi^I)\phi^J
-\frac{1}{z\partial^+}\partial^+\xi^z \phi^I\,.
\end{eqnarray*}
Here and below we use the notation $\xi\partial\equiv \xi^\mu\partial_\mu$.
To simplify our expressions we use creation and annihilation
operators $\alpha^I$, $\bar{\alpha}^I$ (\ref{comosc})
and introduce Fock vector for the physical field $\phi^I$

$$
|\phi\rangle\equiv \phi^I\alpha^I|0\rangle\,.
$$
The transformation laws of the physical field can be then cast into the
following form

\begin{equation}\label{vectra3}
\delta_{tot}|\phi\rangle
=\Bigl(\xi\partial+\frac{d-2}{2z}\xi^z
+\frac{1}{2}\partial^I\xi^JM^{IJ}
+M^{IJ}\partial^+\xi^I\frac{\partial^J}{\partial^+}
-\frac{d-4}{2z\partial^+}\partial^+\xi^IR^{zI}
-\frac{\partial^+\xi^z}{z\partial^+}
\Bigr)|\phi\rangle
\end{equation}
where the spin operator $M^{AB}$ was defined by (\ref{lorspiope}) while
the operator $R^{AB}$ is given by

\begin{equation}\label{rab}
R^{AB}\equiv \alpha^A\bar{\alpha}^B+\alpha^B\bar{\alpha}^A\,.
\end{equation}
The second and last two terms in r.h.s. of (\ref{vectra3}) are absent in
Poincar\'e algebra transformation laws. It is these terms that break manifest
$so(d-2)$ invariance to $so(d-3)$ one.
Note that the operators $M^{IJ}$ and $R^{IJ}$ form $gl(d-2)$ algebra
which has $so(d-2)$ subalgebra spanned by the spin operator $M^{IJ}$.
Appearance of $R^{IJ}$ is not desirable. Fortunately, it turns out  that
the last two terms in r.h.s. of (\ref{vectra3}) can be expressed in terms
of square of $M^{IJ}$. To avoid repetition we will demonstrate this
explicitly when we shall consider arbitrary spin $s$ field whose
particular case is the Maxwell spin one field.

\subsection{Totally antisymmetric fields. Light-cone form of equations
of motion}

The next useful toy model is totally antisymmetric field
$A^{\mu_1\ldots \mu_s}$. As before we prefer to use
tangent space field $\Phi^{A_1\ldots A_s}$ defined by (\ref{tantar}).
Here from the very beginning we start with the generating function
(\ref{genfun}) and use the anticommuting oscillators (\ref{antosc}).
An appropriate generating function for the field strength is given by

$$
|F\rangle=\alpha D|\Phi\rangle\,,
$$
where the notation is given in (\ref{lorspiope}), (\ref{conv}). The gauge
transformation in terms of generating function $|\Phi\rangle$ take the
form

\begin{equation}
\label{asfgtr1}
\delta_{gt}|\Phi\rangle= \alpha D|\Lambda\rangle\,.
\end{equation}
The fact that $|F\rangle$ is indeed invariant with respect of this
transformation can be easily seen from the relation
$(\alpha D)^2=0$. In terms of $|F\rangle$ the equations of motion take
the form

\begin{equation}\label{asyeqmot0}
\bar{\alpha}D|F\rangle=0\,.
\end{equation}
Making use of relations

$$
\alpha D=\alpha\hat{\partial}-\alpha^z \alpha\bar{\alpha}\,,
\qquad
\bar{\alpha} D =\bar{\alpha}\hat{\partial}
-d\bar{\alpha}^z+\bar{\alpha}^z\alpha\bar{\alpha}\,,
$$
we transform the equations of motion (\ref{asyeqmot0}) to the following
form

\begin{equation}\label{asfeqmot2}
(\bar{\alpha}\hat{\partial}+(s+1-d)\bar{\alpha}^z)
(\alpha\hat{\partial}-s\alpha^z)|\Phi\rangle=0\,.
\end{equation}
The invariance with respect to the gauge transformation (\ref{asfgtr1})
allows us to impose light-cone gauge which in terms of generating
function looks as

\begin{equation}
\label{asflcg}
\bar{\alpha}^+|\Phi\rangle=0\,.
\end{equation}
By applying $\bar{\alpha}^+$ to the equations of motion and taking into
account this  gauge we get a constraint which dividing by
$\hat{\partial}^+$ can be cast into  the form

$$
(\bar{\alpha}\hat{\partial}+(s+2-d)\bar{\alpha}^z)|\Phi\rangle=0\,.
$$
Taking into account the gauge (\ref{asflcg}) we get from this the
following constraint

$$
(\bar{\alpha}^-\hat{\partial}^+
+\bar{\alpha}^I\hat{\partial}^I+(s+2-d)\bar{\alpha}^z)|\Phi\rangle=0\,.
$$
Solution to this constraint is given by

\begin{equation}
\label{asfsolcon}
|\Phi\rangle
=\Bigl(1-\frac{\partial^I}{\partial^+}\alpha^+\bar{\alpha}^I
+\frac{d-s-2}{\hat{\partial}^+}\alpha^+\bar{\alpha}^z\Bigr)
|\Phi_{ph}\rangle\,,
\end{equation}
where $|\Phi_{ph}\rangle$ is generating function of physical degrees of
freedom, i.e. it depends only on $\alpha^I$:

$$
|\Phi_{ph}\rangle
=\Phi^{I_1\ldots I_s}\alpha^{I_1}\ldots \alpha^{I_s}|0\rangle\,.
$$
Now by inserting (\ref{asfsolcon}) into equations
of motion (\ref{asfeqmot2}) we get the following
equations of motion for physical field $|\Phi_{ph}\rangle$

\begin{equation}\label{asyeqmot1}
(z^2\partial^2+(2-d)z\partial_z+(d-2s-2)\alpha^z\bar{\alpha}^z
+s(d-s-1))|\Phi_{ph}\rangle=0\,.
\end{equation}
We wish to express oscillator part of these equitations in terms of spin
operator $M^{IJ}$ alone. By using the representation (\ref{lorspiope}) we
get

\begin{eqnarray*} &&
\frac{1}{2}M_{ij}^2|\Phi_{ph}\rangle
=(s(s+3-d)+(d-2s-2)\alpha^z\bar{\alpha}^z)|\Phi_{ph}\rangle\,,
\\
&&
M_{IJ}^2|\Phi_{ph}\rangle
=-2s(d-2-s)|\Phi_{ph}\rangle\,,
\end{eqnarray*}
where
$$
M_{ij}^2\equiv M^{ij}M^{ij}\,,
\qquad
M_{IJ}^2\equiv M^{IJ}M^{IJ}\,.
$$
With the help of these relationships we can cast the equations of motion
(\ref{asyeqmot1}) into the following form

$$
(z^2\partial^2+(2-d)z\partial_z
+\frac{1}{2}M_{ij}^2-M_{IJ}^2)|\Phi_{ph}\rangle=0\,.
$$
Note that $M_{ij}^2$ and $M_{IJ}^2$ are nothing but the second order
Casimir operator of the $so(d-3)$ and $so(d-2)$ algebras respectively. The
fact that we can cast our equations into this form we consider as one of
interesting results of this work. In order to cancel $z\partial_z$ term we
make the rescaling

\begin{equation}
\label{Phiphi}
|\Phi_{ph}\rangle=z^{(d-2)/2}|\phi\rangle
\end{equation}
and in such a way we get desirable light-cone form of equations of motion
for physical field

$$
\Bigl(z^2\partial^2
+\frac{1}{2}M_{ij}^2-M_{IJ}^2
-\frac{d(d-2)}{4}\Bigr)|\phi\rangle=0\,.
$$
By rewriting this equation in the Schr\"odinger form

\begin{equation}\label{schfor2}
\partial^-|\phi\rangle =P^-|\phi\rangle
\end{equation}
we can immediately get the hamiltonian

\begin{equation}\label{asyham}
P^-=-\frac{\partial_I^2}{2\partial^+}
+\frac{1}{2z^2\partial^+}\Bigl(-\frac{1}{2}M_{ij}^2+M_{IJ}^2
+\frac{d(d-2)}{4}\Bigr)\,.
\end{equation}
The action that leads to equations of motion (\ref{schfor2}) looks as

\begin{equation}\label{lcaction}
S_{l.c.}
= \int d^dx \langle \partial^+\phi|(-\partial^- +P^-)|\phi\rangle\,.
\end{equation}
It is instructive to demonstrate how do the results of this section
reproduce the ones for the spin one Maxwell field.
Spin one vector field $\phi^I$, which transforms in vector representation
of $so(d-2)$ algebra, is decomposed into vector representation $\phi^i$ and
scalar representation $\phi$ of $so(d-3)$ algebra. For both these
components the Casimir operator of $so(d-2)$ algebra $M_{IJ}^2$ takes the
same values\footnote{Here we use the fact that for totally antisymmetric
spin $s$ representation of $so(N)$ algebra the Casimir operator $M_{IJ}^2$
takes the value $M_{IJ}^2|\phi\rangle=-2s(N-s)|\phi\rangle$.}

$$
M_{IJ}^2|\phi\rangle=-2(d-3)|\phi\rangle\,,
\qquad
|\phi\rangle=|\phi_1\rangle+|\phi_0\rangle\,,
\quad
|\phi_1\rangle=\phi^i\alpha^i|0\rangle\,,
\quad
|\phi_0\rangle=\phi\alpha^z|0\rangle
$$
while the Casimir operator of $so(d-3)$ subalgebra $M_{ij}^2$ gives

$$
M_{ij}^2|\phi_1\rangle=-2(d-4)|\phi_1\rangle\,,
\qquad
M_{ij}^2|\phi_0\rangle=0\,.
$$
Now it is straightforward to see that the formulas for
spin one (\ref{hamvec1}),(\ref{hamvec2}) are indeed
reproduced.

\subsection{Light-cone form of transformations of totally
antisymmetric field}

Now we are studying transformation laws of physical degrees of freedom
collected in $|\phi\rangle$.
As in the case spin one field we start with
original global  AdS symmetries, supplemented by
compensating gauge transformation required to maintain the gauge.
The original global AdS transformations in terms of target space tensor
field are given by

\begin{equation}\label{lieder0}
\delta_{isom} A^{\mu_1\ldots\mu_s}
={\cal L}_\xi A^{\mu_1\ldots\mu_s}\,,
\end{equation}
where the action of Lie derivative ${\cal L}_\xi$ is given by
$$
{\cal L}_\xi A^{\mu_1\ldots\mu_s}
=\xi^\nu\partial_\nu A^{\mu_1\ldots\mu_s} -\sum_{k=1}^s
A^{\mu_1\ldots\mu_{k-1}\nu\mu_{k+1}\ldots\mu_s}\partial_\nu
\xi^{\mu_k}\,.
$$
In terms of generating function for the tangent space field
$|\Phi\rangle$ (\ref{genfun})
and tangent space Killing vectors $\eta^A$ (\ref{etxi})
these transformations take the form

\begin{equation}
\label{asftra0}
\delta_{isom}|\Phi\rangle = {\cal L}_\eta|\Phi\rangle\,,
\qquad
{\cal L}_\eta=\eta^AD_A+\frac{1}{2}D^A\eta^B M^{AB}\,,
\end{equation}
In Poincar\'e coordinates the Lie derivative takes the form

\begin{equation}
\label{lieder1}
{\cal L}_\eta
=\eta\hat{\partial}+\frac{1}{2}\hat{\partial}^A\eta^BM^{AB}
+\frac{1}{2}M^{zB}\eta^B\,.
\end{equation}
Now let us focus on the original AdS algebra transformations
supplemented by compensating gauge transformation

\begin{equation}
\label{asftra1}
\delta_{tot}|\Phi\rangle = {\cal L}_\eta|\Phi\rangle
+\alpha D|\Lambda\rangle\,.
\end{equation}
Our aim in is to get transformation laws
for the physical field $|\phi\rangle$ whose relationship to
the gauge field $|\Phi\rangle$ is described by (\ref{asfsolcon})
(\ref{Phiphi}).  First we have to find appropriate $|\Lambda\rangle$. To
this end we cast the transformations (\ref{asftra1}) into the following
form

\begin{equation}
\label{asftra2}
\delta_{tot}|\Phi\rangle =
(\eta\hat{\partial}+\frac{1}{2}\hat{\partial}^A\eta^BM^{AB}
+\frac{1}{2}M^{zB}\eta^B)|\Phi\rangle
+(\alpha\hat{\partial}-(s-1)\alpha^z)|\Lambda\rangle\,.
\end{equation}
Then from the relation

\begin{equation}\label{lamcon}
\bar{\alpha}^+ \delta_{tot}|\Phi\rangle=0\,,
\end{equation}
which is amount to the requirement the complete transformations
(\ref{asftra1}) maintain
the gauge (\ref{asflcg}) we find the equation

$$
\hat{\partial}^+ \eta^I\bar{\alpha}^I|\Phi\rangle
+(\hat{\partial}^++\alpha^z\bar{\alpha}^+
-\alpha D\bar{\alpha}^+)|\Lambda\rangle=0\,,
$$
which obviously has the following solution

\begin{equation}
\label{Lamsol}
|\Lambda\rangle
=-\frac{\partial^+\eta^I}{\partial^+}\bar{\alpha}^I|\Phi\rangle\,.
\end{equation}
Note that this solution is fixed by module the term
$\alpha D|f\rangle$. Because this term  does obviously not contribute to
transformation of $|\Phi\rangle$ we interested in we can ignore it.

Second let us note that in light-cone gauge the physical field
$|\Phi_{ph}\rangle$ is a coefficient $(\alpha^+)^0$ in $|\Phi\rangle$. Due
to that in order to find contribution to $\delta_{tot}|\Phi_{ph}\rangle$ it
is sufficient to analyse $(\alpha^\pm)^0$ terms in (\ref{asftra1}).
The light-cone gauge and the requirement (\ref{lamcon}) which is already
satisfied tell us that the terms proportional to $\alpha^-$ are absent
in both sides of (\ref{asftra1}).  Therefore we can restrict our attention
to $(\alpha^+)^0$ terms in (\ref{asftra1}).  In this way we get the
following extremely useful formula for Lie derivative of physical field

\begin{equation}\label{ldpf}
{\cal L}_\eta|\Phi_{ph}\rangle
=(\eta\hat{\partial}
+\frac{1}{2}\hat{\partial}^I\eta^JM^{IJ}
+\frac{1}{2}\eta^IM^{zI})|\Phi_{ph}\rangle
-\hat{\partial}^+\eta^I\alpha^I
(\bar{\alpha}^-|\Phi\rangle)\Bigr|_{\alpha^+=0}^{\vphantom{5pt}}\,,
\end{equation}
where equality $\alpha^+=0$ indicates that we consider only $(\alpha^+)^0$
terms. The last term in (\ref{ldpf}) can immediately be evaluated
by using (\ref{asfsolcon})

\begin{equation}\label{amphi}
\bar{\alpha}^-|\Phi\rangle
=(-\frac{\partial^I}{\partial^+}\bar{\alpha}^I
+\frac{d-s-2}{\hat{\partial}^+}\bar{\alpha}^z)|\Phi_{ph}\rangle\,.
\end{equation}
Thus transformation laws of physical field are given by

\begin{equation}\label{phtr}
\delta_{tot}|\Phi_{ph}\rangle
={\cal L}_\eta|\Phi_{ph}\rangle+
(\alpha^I\hat{\partial}^I-(s-1)\alpha^z)|\Lambda\rangle
\Bigr|_{\alpha^+=0}^{\vphantom{5pt}}\,.
\end{equation} From (\ref{Lamsol})
and (\ref{asfsolcon}) we get immediately

$$
|\Lambda\rangle|_{\alpha^+=0}^{\vphantom{5pt}}
=-\frac{\partial^+\eta^I}{\partial^+}\bar{\alpha}^I|\Phi_{ph}\rangle\,.
$$
Inserting this into (\ref{phtr}) and taking into account
(\ref{ldpf}),(\ref{amphi}) we get the following transformation laws for
physical field

\begin{eqnarray*}
\delta_{tot}|\Phi_{ph}\rangle
&=&\Bigl(\eta\hat{\partial}
+\frac{1}{2}\hat{\partial}^I\eta^J M^{IJ}
+\frac{1}{2}\eta^IM^{zI}
+M^{IJ}\hat{\partial}^+\eta^I\frac{\partial^J}{\partial^+}
+\frac{d-2}{2}M^{zI}\frac{\partial^+\eta^I}{\partial^+}
\\
&+&\frac{2s+2-d}{2\partial^+}\partial^+\eta^IR^{zI}
-s\frac{\partial^+\eta^z}{\partial^+}\Bigr)|\Phi_{ph}\rangle\,.
\end{eqnarray*}
In terms of physical field $|\phi\rangle$ (\ref{Phiphi}) and target space
Killing vectors $\xi^\mu$ (\ref{etxi}) this transforms to

\begin{equation}\label{astraint}
\delta_{tot}|\phi\rangle
=\Bigl(\xi\partial+\frac{d-2}{2z}\xi^z
+\frac{1}{2}\partial^I\xi^JM^{IJ}
+M^{IJ}\partial^+\xi^I\frac{\partial^J}{\partial^+}
+\frac{2s+2-d}{2z\partial^+}\partial^+\xi^IR^{zI}
-s\frac{\partial^+\xi^z}{z\partial^+}\Bigr)|\phi\rangle
\end{equation}
Note that for spin one this expression reduces to the one in
(\ref{vectra3}). As before one has term proportional to $R^{zI}$. Now let
us demonstrate that this terms can be entirely expressed in terms of the
spin operator $M^{IJ}$. Indeed making of the following relationships

$$
\{M^{zj},M^{jI}\}=(d-2-2\alpha^J\bar{\alpha}^J)R^{zI}
+2\delta^I_z\alpha^J\bar{\alpha}^J
$$
it is straightforward to see that the transformations (\ref{astraint}) can
be cast into  the following desired form

\begin{equation}\label{astrafin}
\delta_{tot}|\phi\rangle
=\Bigl(\xi\partial+\frac{d-2}{2z}\xi^z
+\frac{1}{2}\partial^I\xi^JM^{IJ}
+M^{IJ}\partial^+\xi^I\frac{\partial^J}{\partial^+}
-\frac{\partial^+\xi^I}{2z\partial^+}\{M^{zj},M^{jI}\}
\Bigr)|\phi\rangle\,.
\end{equation}

This remarkable formula demonstrates that
{\it the light-cone transformations
of physical fields are indeed expressible in terms of the spin operator}
$M^{IJ}$. Note that Poincar\'e algebra transformations involve   only
terms linear in $M^{IJ}$. The presence of the term {\it quadratic}
in $M^{IJ}$
is an  essential feature of the AdS algebra transformations. At the same
time the AdS transformations do not involve higher than second order terms
in $M^{IJ}$.  This implies, at least, that light-cone formalism in AdS
space-time should not be  much more complicated than the one in Minkowski
space-time. With this optimistic conclusion let us proceed to the totally
symmetric fields.

\subsection{Totally symmetric fields in AdS space-time. Gauge invariant
equations of motion}

The case of totally symmetric fields is the most interesting one
because of the
following reasons. One is that the graviton falls into this representation.
Another is that  in four dimensional AdS space-time all massless physical
fields are described by totally symmetric fields. Some time ago completely
self consistent interacting equations of motion for such fields have been
discovered \cite{vas1}.
Therefore,  before
 proceeding  to the  general light-cone case
 it is reasonable to study the totally
symmetric fields in their own right.  Note that we consider, as
before, an  arbitrary $d$-dimensional AdS space-time.
The starting point in our derivation of light-cone equations of
motion are the
gauge invariant equations of motion.\footnote{Gauge
invariant action in $d=4$ AdS space-time for integer spin field
has been established in \cite{F1} while for half integer spin in
\cite{FF}. We do not rely on these results because we need equations of
motion for arbitrary space-time dimension. For arbitrary space-time
dimension the gauge invariant action for massless field has been found
in (\cite{vas5},\cite{vas6}). There the action has been constructed in
terms of linear field strength and the reason because we do not use these
results is that it is quite difficult to transform the equations of
motion for field strengths to the ones for gauge fields. The equations of
motion for arbitrary space-time dimension in terms of gauge fields have been
found in (\cite{metsit1}-\cite{metsit6}) but there these equations have
been formulated in Lorentz gauge for gauge fields. However as we already
demonstrated in light-cone gauge the Lorentz constraint does not follow
from the AdS gauge invariant equations of motion.  Thus for the purposes
of this work we need to derive the equations in question from the very
beginning.} Some details of derivation of these equations of motion can
be found in Appendix A. Let us present the result.

We formulate our equations of motion in terms of the generating function
$|\Phi\rangle$ defined by (\ref{genfun}) where $\Phi^{A_1\ldots A_s}$ is
totally symmetric tangent space tensor (\ref{tantar}) while $\alpha^A$,
$\bar{\alpha}^A$ are commuting oscillators defined by  (\ref{comosc}).
The fact that  $|\Phi\rangle$ is a spin $s$ field is reflected by
constraint

$$
\alpha\bar{\alpha}|\Phi\rangle=s|\Phi\rangle\,,
\quad
\alpha\bar{\alpha}\equiv \alpha^A\bar{\alpha}^A\,.
$$
In addition one imposes the double traceless condition

$$
(\bar{\alpha}^2)^2|\Phi\rangle=0\,,
\qquad
\bar{\alpha}^2\equiv \bar{\alpha}^A\bar{\alpha}^A\,.
$$
The gauge invariant equations of motion look then as

\begin{equation}
\label{symeqmot1}
\Bigl([\bar{\alpha}D,\alpha D]
-\alpha D\bar{\alpha}D+\frac{1}{2}(\alpha D)^2\bar{\alpha}^2
-2\alpha^2\bar{\alpha}^2+2(2s+d-3)\Bigr)|\Phi\rangle=0\,,
\end{equation}
where the corresponding gauge transformation is

\begin{equation}\label{dfa1}
\delta |\Phi\rangle=\alpha D|\Lambda\rangle
\end{equation}
and the gauge parameter field $\Lambda$ is subject to usual
traceless condition

\begin{equation}\label{lamtracon1}
\bar{\alpha}^2|\Lambda\rangle=0\,.
\end{equation}
The above equations of motion can equivalently be rewritten in the
form

\begin{equation}
\label{symeqmot1.0}
(D_A^2+\omega^{AAB}D_B-\alpha D\bar{\alpha}D
+\frac{1}{2}(\alpha D)^2\bar{\alpha}^2
-\alpha^2\bar{\alpha}^2-s^2+(6-d)s+2d-6)|\Phi\rangle=0\,,
\end{equation}
which is more convenient in practical calculations.
The formulation in terms of double traceless field is not convenient
for our purposes. As is well known the double traceless spin $s$ field
can be decomposed into spin $s$ traceless field the one of spin $s-2$:

\begin{equation}\label{dec}
|\Phi\rangle=|\Phi_s\rangle+\alpha^2|\Phi_{s-2}\rangle\,,
\end{equation}
where the fields
$|\Phi_s\rangle$ and $|\Phi_{s-2}\rangle$ satisfy the usual traceless
condition

\begin{equation}\label{trcos}
\bar{\alpha}^2|\Phi_s\rangle=0\,,
\qquad
\bar{\alpha}^2|\Phi_{s-2}\rangle=0\,.
\end{equation}
Inverse relations to the decomposition (\ref{dec}) are
$$
|\Phi_s\rangle
=(1-\frac{\alpha^2\bar{\alpha}^2}{2(2s+d-4)})|\Phi\rangle\,,
\qquad
2(2s+d-4)|\Phi_{s-2}\rangle
=\bar{\alpha}^2|\Phi\rangle\,.
$$
In terms of these new fields the equations of motion
(\ref{symeqmot1.0}) take the form

\begin{eqnarray}
&&
\Bigl(D_A^2+\omega^{AAB}D_B-\alpha D\bar{\alpha}D
-s^2+(6-d)s+2d-6\Bigr)|\Phi_s\rangle
\nonumber\\
&&
+\alpha^2\Bigl(D_A^2+\omega^{AAB}D_B-\alpha D\bar{\alpha}D
-s^2+(2-d)s+2\Bigr)|\Phi_{s-2}\rangle
\nonumber\\
\label{symeqmot2}
&&+(2s+d-6)(\alpha D)^2|\Phi_{s-2}\rangle=0\,.
\end{eqnarray}
This form  turns out to be more convenient for deriving light-cone form
of equations of motion. Gauge transformation for $|\Phi_s\rangle$ looks as

\begin{equation}\label{gts}
\delta|\Phi_s\rangle=\Bigl(\alpha D
-\frac{\alpha^2}{2s+d-4}\bar{\alpha}D\Bigr)|\Lambda\rangle\,.
\end{equation}
Gauge transformation for the field $|\Phi_{s-2}\rangle$ can be
obtained straightforwardly  from the formulas above but we do not need
them because in light-cone gauge the constraints produced by equations of
motion set the field $|\Phi_{s-2}\rangle$ to be equal zero.

\subsection{Totally symmetric fields. Light-cone form of
equations of motion}

Taking into account that the gauge field $|\Phi_s\rangle$ and the gauge
parameter field $|\Lambda\rangle$ have the same number degrees of freedom
and due to invariance with respect to gauge transformation (\ref{gts})
we can impose the light-cone gauge

\begin{equation}\label{lcs}
\bar{\alpha}^+|\Phi_s\rangle=0\,.
\end{equation}
Note that in this gauge we get from (\ref{trcos}) the constraint

\begin{equation}\label{trcon1}
\bar{\alpha}_I^2|\Phi_s\rangle=0\,,
\end{equation}
i.e. $|\Phi_s\rangle$ becomes traceless field with respect to
transverse indices. Acting with $\bar{\alpha}^{+2}$ on the equations
of motion (\ref{symeqmot2}) one proves that $|\Phi_{s-2}\rangle=0$. Then by
acting $\bar{\alpha}^+$ on the equations of motion we get the Lorentz
like constraint

$$
\bar{\alpha}D|\Phi_s\rangle
=-\frac{2(\hat{\partial}^+
-\alpha^+\bar{\alpha}^z)}{\hat{\partial}^+-2\alpha^+\bar{\alpha}^z}
\bar{\alpha}^z|\Phi_s\rangle\,.
$$
Dividing both the sides by $\partial^+$ this constraint can be cast into
form

\begin{equation}\label{symlorcon}
\bar{\alpha}^-|\Phi_s\rangle
=\Bigl(-\frac{\partial^I}{\partial^+}\bar{\alpha}^I
+\frac{s+d-2}{\hat{\partial}^+}\bar{\alpha}^z
-\frac{2(\hat{\partial}^+
-\alpha^+\bar{\alpha}^z)}{\hat{\partial}^+(\hat{\partial}^+
-2\alpha^+\bar{\alpha}^z)}\bar{\alpha}^z\Bigr)
|\Phi_s\rangle\,.
\end{equation}
Solution to this equation is found to be

\begin{equation}
\label{symlcsol}
|\Phi_s\rangle=P|\Phi_{ph}\rangle\,,
\end{equation}
where an operator $P$ is defined by

$$
P\equiv \Bigl(1-\frac{2}{\hat{\partial}^+}\alpha^+\bar{\alpha}^z\Bigr)
\Bigl(1-\frac{1}{\hat{\partial}^+}\alpha^+\bar{\alpha}^z\Bigr)^{2-d-s}
\exp(-\frac{\partial^I}{\partial^+}\alpha^+\bar{\alpha}^I)\,.
$$
The generating function $|\Phi_{ph}\rangle$ consists of only physical
degrees of freedom

$$
|\Phi_{ph}\rangle=\Phi^{I_1\ldots
I_s}\alpha^{I_1}\ldots \alpha^{I_s}|0\rangle
$$
and due to (\ref{trcon1}) fulfills the traceless condition

\begin{equation}\label{trcon2}
\bar{\alpha}_I^2|\Phi_{ph}\rangle=0\, ,
\end{equation}
which tells us  that
the field $|\Phi_{ph}\rangle$ has the number of spin degrees
of freedom equal to dimension of symmetric spin $s$ irreducible
representation of $so(d-2)$ algebra. It is the field $|\Phi_{ph}\rangle$
that describes physical degrees of freedom. Now we have to insert solution
for $|\Phi_s\rangle$ (\ref{symlcsol}) into equations of motion
(\ref{symeqmot2}). Taking into account that
$|\Phi_{s-2}\rangle=0$ and rewriting the equations of motion
(\ref{symeqmot2}) explicitly in Poincar\'e coordinates we get

$$
\Bigl((\hat{\partial}^A-\alpha^A\bar{\alpha}^z)^2
+(2\alpha^z-\alpha D)\bar{\alpha}D
+(1-d)(\hat{\partial}^z-\alpha^z\bar{\alpha}^z)
-s^2+(5-d)s+2d-6\Bigr)|\Phi_s\rangle=0\,.
$$
Inserting the solution (\ref{symlcsol}) in these equations we find
after some tedious but straightforward calculations the following
equations of motion for the physical field $|\Phi_{ph}\rangle$

$$
\Bigl(\hat{\partial}^2+(1-d)\hat{\partial}^z
+(2s+d-6)\alpha^z\bar{\alpha}^z-\alpha_I^2\bar{\alpha}^{z2}
-s^2+(5-d)s+2d-6\Bigr)|\Phi_{ph}\rangle=0\,.
$$
As before we wish to express the oscillator part of these equations
in terms of spin operator $M^{IJ}$ alone. Taking into account
the relation

$$
\frac{1}{2}M_{ij}^2=\alpha_i^2\bar{\alpha}_j^2
-(\alpha^j\bar{\alpha}^j)^2+(5-d)\alpha^j\bar{\alpha}^j
$$
and the traceless condition (\ref{trcon2}) one easily
proves the following relationship

$$
\frac{1}{2}M_{ij}^2|\Phi_{ph}\rangle
=\Bigl(-\alpha_J^2\bar{\alpha}^{z2}
+(2s+d-6)\alpha^z\bar{\alpha}^z-s^2+(5-d)s
\Bigr)|\Phi_{ph}\rangle\,.
$$
Making use of this relationship we get the
following nice representation for the {\it light-cone equations of motion}

$$
\Bigl(\hat{\partial}^2+(1-d)\hat{\partial}^z
+\frac{1}{2}M_{ij}^2+2d-6\Bigr)|\Phi_{ph}\rangle=0\,.
$$
Finally, in terms of the canonical normalized physical field $|\phi\rangle$
defined as in (\ref{Phiphi}) one has the following form for equations of
motion

\begin{equation}\label{symeqmot5}
\Bigl(z^2\partial^2+\frac{1}{2}M_{ij}^2
-\frac{(d-4)(d-6)}{4}\Bigr)|\phi\rangle=0\,.
\end{equation}
Transforming this equations into  the Schr\"odinger form
(\ref{schfor2}) we get the generator $P^-$ for physical totally symmetric
field

\begin{equation}\label{symham}
P^-=-\frac{\partial_I^2}{2\partial^+}
+\frac{1}{2z^2\partial^+}(-\frac{1}{2}M_{ij}^2+\frac{(d-4)(d-6)}{4})\,.
\end{equation}
By inserting this $P^-$ and the field $|\phi\rangle$ into (\ref{lcaction})
we get immediately the action that leads to the above light-cone equations
of motion.

\subsection{Light-cone form of transformations of totally
symmetric field}

In order to find transformation laws of the physical field
$|\phi\rangle$ we start as before
with the global transformations for the gauge field $|\Phi_s\rangle$
supplemented by appropriate compensating gauge transformation

\begin{equation}\label{totrs}
\delta_{tot}|\Phi_s\rangle
={\cal L}_\eta|\Phi_s\rangle+\delta_{gt}|\Phi_s\rangle\,,
\end{equation}
where the gauge transformation is given in (\ref{gts}).
The Lie derivative is given in
(\ref{lieder1}), where the anticommuting oscillators should
be replaced by commuting ones. As above the gauge
parameter field $|\Lambda\rangle$ is found from the equation

\begin{equation}\label{symlamsol}
\bar{\alpha}^+\delta_{tot}|\Phi_s\rangle=0\,.
\end{equation}
Note that in order to find
$\delta_{tot}|\Phi_{ph}\rangle$ it is sufficient to analyse
$(\alpha^\pm)^0$ terms in (\ref{totrs}).
Before to solve the equation (\ref{symlamsol}) we consider the
contributions of global and gauge transformations in turn.
Let us first consider original global AdS
transformations. To this end we can use the formula (\ref{ldpf}). The last
term in (\ref{ldpf}) can be obtained from (\ref{symlorcon}) and is given by

\begin{equation}\label{symampf}
\bar{\alpha}^-|\Phi_s\rangle|_{\alpha^+=0}^{\vphantom{5pt}}
=(-\frac{\partial^I}{\partial^+}\bar{\alpha}^I
+\frac{s+d-4}{\hat{\partial}^+}\bar{\alpha}^z)|\Phi_{ph}\rangle\,.
\end{equation}
Inserting this into (\ref{ldpf}) we get the following
contribution of the original global AdS transformations

\begin{equation}\label{LiPhs}
{\cal L}_\eta|\Phi_{ph}\rangle
=\Bigl(\eta\hat{\partial}
+\frac{1}{2}\hat{\partial}^I\eta^JM^{IJ}
+\frac{1}{2}\eta^IM^{zI}
+\hat{\partial}^+\eta^I\alpha^I
(\frac{\partial^J}{\partial^+}\bar{\alpha}^J
-\frac{s+d-4}{\hat{\partial}^+}\bar{\alpha}^z)
\Bigr)|\Phi_{ph}\rangle\,.
\end{equation}
Now we should find a contribution of the gauge transformation.
To this end we cast the gauge transformation
(\ref{gts}) into the form

\begin{equation}\label{gts2}
\delta_{gt}|\Phi_s\rangle
=\Bigl(\alpha\hat{\partial}+(s-1)\alpha^z
-\frac{\alpha^2}{2s+d-4}(\bar{\alpha}\hat{\partial}
+(s-1)\bar{\alpha}^z)\Bigr)|\Lambda\rangle\,.
\end{equation}
Since in r.h.s. of this expressions there is the annihilation
oscillator $\bar{\alpha}^-$ it is clear that in order to find contribution
of $|\Lambda\rangle$ into $\delta_{tot}|\Phi_{ph}\rangle$ we need only the
first two leading terms in expansion of $|\Lambda\rangle$ in powers of
$\alpha^+$:

$$
|\Lambda\rangle=|\Lambda_0\rangle
+\alpha^+|\Lambda_1\rangle+\ldots\,.
$$
Plugging this into (\ref{gts2}) and taking $(\alpha^\pm)^0$
terms we find

\begin{equation}
\label{gts3}
\delta_{gt}|\Phi_{ph}\rangle
=(\alpha^I\hat{\partial}^I+(s-1)\alpha^z)|\Lambda_0\rangle
-\frac{\alpha_J^2}{2s+d-4}
\Bigl((\bar{\alpha}^I\hat{\partial}^I+(s-1)\bar{\alpha}^z)
|\Lambda_0\rangle
+\hat{\partial}^+|\Lambda_1\rangle\Bigr)\,.
\end{equation}
All that remains to find explicit transformation is to
find solution to $|\Lambda\rangle$. To this end we return to the
equation (\ref{symlamsol}). Using there an explicit form of the Lie
derivative (\ref{lieder1}) and the gauge transformation (\ref{gts}) we get
from (\ref{symlamsol}) the following equation for $|\Lambda\rangle$

\begin{equation}\label{figpeq}
\hat{\partial}^+\eta^I\bar{\alpha}^I|\Phi_s\rangle
+\Bigl(\hat{\partial}^+
-\frac{2\alpha^+}{2s+d-4}(\bar{\alpha}\hat{\partial}
+(s-1)\bar{\alpha}^z)\Bigr)|\Lambda\rangle=0\,.
\end{equation}
Note that while deriving this equation we have used
the fact that $\bar{\alpha}^+|\Lambda\rangle=0$. This fact can be proved
by analysis the requirement $\bar{\alpha}^{+2}\delta_{tot}|\Phi_s\rangle=0$
which is consequence of (\ref{symlamsol}).  From the equation
(\ref{figpeq}) we get immediately

\begin{equation}\label{Laze}
|\Lambda_0\rangle
=-\frac{\partial^+\eta^I}{\partial^+}\bar{\alpha}^I|\Phi_{ph}\rangle\,.
\end{equation}
Then by acting on (\ref{figpeq}) with $\bar{\alpha}^-$ and taking
$(\alpha^+)^0$ terms we obtain the relation

$$
\hat{\partial}^+\eta^I\bar{\alpha}^I\bar{\alpha}^-|\Phi_s\rangle
+\Bigl(\bar{\alpha}^-\hat{\partial}^+
-\frac{2}{2s+d-4}(\bar{\alpha}^-\hat{\partial}^+
+\bar{\alpha}^I\hat{\partial}^I
+(s-1)\bar{\alpha}^z)\Bigr)|\Lambda\rangle
\Bigr|_{\alpha^+=0}^{\vphantom{5pt}}=0\,.
$$
Taking into account (\ref{symampf}) we get

\begin{equation}\label{Lafi}
(\bar{\alpha}^I\hat{\partial}^I
+(s-1)\bar{\alpha}^z)|\Lambda_0\rangle
+\hat{\partial}^+|\Lambda_1\rangle
=-(2s+d-4)\frac{\partial^+\eta^I}{\partial^+}
\bar{\alpha}^I\bar{\alpha}^z|\Phi_{ph}\rangle\,.
\end{equation}
By plugging (\ref{Laze}) and (\ref{Lafi}) into (\ref{gts3})
and taking into account the relation

$$
(\alpha^I\hat{\partial}^I+(s-1)\alpha^z)|\Lambda_0\rangle
=\Bigl(-\hat{\partial}^+\eta^J(\alpha^I\frac{\partial^I}{\partial^+}
+\frac{s-2}{\hat{\partial}^+}\alpha^z)\bar{\alpha}^J
-s\frac{\partial^+\eta^z}{\partial^+}\Bigr)|\Phi_{ph}\rangle\,.
$$
we get contribution of the gauge transformation
$$
\delta_{gt}|\Phi_{ph}\rangle
=\Bigl(-\hat{\partial}^+\eta^J(\alpha^I\frac{\partial^I}{\partial^+}
+\frac{s-2}{\hat{\partial}^+}\alpha^z)\bar{\alpha}^J
+\frac{\partial^+\eta^I}{\partial^+}\alpha_J^2
\bar{\alpha}^I\bar{\alpha}^z
-s\frac{\partial^+\eta^z}{\partial^+}
\Bigr)|\Phi_{ph}\rangle\,.
$$
By summing this with contribution of the original global AdS
transformations (\ref{LiPhs}) we get complete transformation laws for
the physical filed $|\Phi_{ph}\rangle$:

\begin{eqnarray*}
\delta_{tot}|\Phi_{ph}\rangle
&=&\Bigl(\eta\hat{\partial}+\frac{1}{2}\hat{\partial}^I\eta^JM^{IJ}
+\frac{1}{2}\eta^IM^{zI}+M^{IJ}\hat{\partial}^+\eta^I
\frac{\partial^J}{\partial^+}
+\frac{d-2}{2\partial^+}\partial^+\eta^IM^{zI}
\\
&-&\frac{2s+d-6}{2\partial^+}\partial^+\eta^IR^{zI}
+\frac{\partial^+\eta^I}{\partial^+}\alpha_J^2
\bar{\alpha}^I\bar{\alpha}^z
-s\frac{\partial^+\eta^z}{\partial^+}
\Bigr)|\Phi_{ph}\rangle\,.
\end{eqnarray*}
This can be rewritten in terms of the canonical normalized
physical field $|\phi\rangle$ (\ref{Phiphi}) and target Killing vectors
$\xi^\mu$ (\ref{etxi}) as follows

\begin{eqnarray}
\delta_{tot}|\phi\rangle
&=&\Bigl(\xi\partial+\frac{d-2}{2z}\xi^z
+\frac{1}{2}\partial^I\xi^JM^{IJ}
+M^{IJ}\partial^+\xi^I\frac{\partial^J}{\partial^+}
\nonumber\\
\label{symtraint}
&-&\frac{2s+d-6}{2z\partial^+}\partial^+\xi^IR^{zI}
+\frac{\partial^+\xi^I}{z\partial^+}\alpha_J^2
\bar{\alpha}^I\bar{\alpha}^z
-s\frac{\partial^+\xi^z}{z\partial^+}
\Bigr)|\phi\rangle\,.
\end{eqnarray}
This form of transformations is quite different  from the analogous
expressions
for antisymmetric field (\ref{astraint}). Despite this fact,
it turns out that the AdS transformations for totally symmetric physical
field  can also be entirely expressed in terms of the spin operator
$M^{IJ}$.  Indeed,  making use of the  relation

$$
\{M^{zj},M^{jI}\}
=(2\alpha^J\bar{\alpha}^J+d-6)R^{zI}
-2\alpha_J^2\bar{\alpha}^I\bar{\alpha}^z
-2\alpha^I\alpha^z\bar{\alpha}^{J2}
+2\delta_z^I\alpha^J\bar{\alpha}^J
$$
it is straightforward to see that (\ref{symtraint}) can be cast into  the
form

\begin{equation}\label{symtrafin}
\delta_{tot}|\phi\rangle
=\Bigl(\xi\partial+\frac{d-2}{2z}\xi^z
+\frac{1}{2}\partial^I\xi^JM^{IJ}
+M^{IJ}\partial^+\xi^I\frac{\partial^J}{\partial^+}
-\frac{\partial^+\xi^I}{2z\partial^+}\{M^{zj},M^{jI}\}
\Bigr)|\phi\rangle\,.
\end{equation}
Thus, as before, light-cone transformations of physical field

(i)  are expressible in terms of the spin operator $M^{IJ}$;

(ii) consist only  of the terms linear and quadratic in $M^{IJ}$, i.e.
do not involve higher powers in $M^{IJ}$.

Note that the form of transformations for totally symmetric
field (\ref{symtrafin}) coincides with the one for totally antisymmetric
field (\ref{astrafin}). Unfortunately, this fact does not imply that
the form of transformations given in (\ref{symtrafin}) and supplied by
appropriate spin operator $M^{IJ}$ is valid for fields of arbitrary
symmetry.  In this respect the situation differs from the one for Poincar\'e
algebra transformations.  The reason for this is that the AdS algebra
transformations involve term quadratic in $M^{IJ}$.  Light-cone form
description of AdS algebra transformation laws for fields of arbitrary
symmetry is given in section 5.

\subsection{Light-cone form of generators of AdS algebra}

Making use of the AdS transformations given in (\ref{astrafin}) and
(\ref{symtrafin}) we can represent them as differential operators acting
on the physical massless field $|\phi\rangle$. Plugging the Killing vectors
(\ref{kilvec}) in transformation laws (\ref{astrafin}) and
(\ref{symtrafin}) we get corresponding differential form of generators.
Let us present the result.

Light-cone form of AdS algebra kinematical generators is given by

\begin{eqnarray}
\label{3spi}
&&
P^i=\partial^i\,,
\\
\label{3spp}
&&
P^+=\partial^+\,,
\\
&&
D=x^+ P^-+x^-\partial^++x^I\partial^I+\frac{d-2}{2}\,,
\\
\label{3sjpm}
&&
J^{+-}=x^+P^--x^-\partial^+\,,
\\
\label{3sjpi}
&&
J^{+i}=x^+\partial^i-x^i\partial^+\,,
\\
&&
J^{ij}=x^i\partial^j-x^j\partial^i+M^{ij}\,,
\\
\label{3skp}
&&
K^+=-\frac{1}{2}(2x^+x^-+x_I^2)\partial^++x^+ D\,,
\\
&&
K^i=-\frac{1}{2}(2x^+x^-+x_J^2)\partial^i+x^i D+M^{iI}x^I+M^{i-}x^+\,,
\end{eqnarray}
where

$$
M^{-i}\equiv M^{iJ}\frac{\partial^J}{\partial^+}
-\frac{1}{2z\partial^+}\{M^{zj},M^{ji}\}\,.
$$
Remaining generators which we refer to  as dynamical generators are given
by

\begin{eqnarray}
&&
P^-=-\frac{\partial_I^2}{2\partial^+}
+\frac{1}{2z^2\partial^+}A\,,
\\
\label{3sjmi}
&&
J^{-i}
=x^-\partial^i-x^i P^-+M^{iJ}\frac{\partial^J}{\partial^+}
-\frac{1}{2z\partial^+}\{M^{zj},M^{ji}\}\,,
\\
\label{3skm}
&&
K^-=-\frac{1}{2}(2x^+ x^-+x_I^2) P^- +x^- D
+ \frac{1}{\partial^+}x^I\partial^J M^{IJ}
-\frac{x^I}{2z\partial^+}\{M^{zJ},M^{JI}\}\,.
\end{eqnarray}
The form of an operator $A$ depends on representation.
For the case of totally antisymmetric massless field the operator $A$
takes the form (see (\ref{asyham}))

$$
A=-\frac{1}{2}M_{ij}^2+M_{IJ}^2+\frac{d(d-2)}{4}\,,
$$
while for totally symmetric massless field one has
(see (\ref{symham}))

$$
A=-\frac{1}{2}M_{ij}^2+\frac{(d-4)(d-6)}{4}\,.
$$
In what follows  the operator $A$ is referred to as AdS mass operator.

A few comments are in order.

(i) the AdS mass operator $A$ for massless
fields does not equal to zero in general. The operator $A$ is equal
to zero only for massless representations which can be realized as
irreducible representations of conformal algebra \cite{metsit3} which for
the case of $d$-dimensional AdS space-time is the $so(d,2)$ algebra.

(ii) Above representations  have been
derived by using the oscillator form realization of the spin operator
$M^{IJ}$ (see (\ref{lorspiope})).  However,  having expressed generators in
terms of the spin operator alone we can use arbitrary form of realization
for the spin operator $M^{IJ}$.  While exploiting such an  arbitrary form
of realization one should keep in mind however that spin operators $M^{IJ}$
for totally symmetric and antisymmetric representations satisfy the
following defining constraints

\begin{equation}\label{defasm}
(M^3)^{[I|J]}
=\frac{1}{2}(M^2+d^2-5d+8)M^{IJ}
\end{equation}
for case of totally antisymmetric field and
\begin{equation}\label{defsymm}
(M^3)^{[I|J]}=(-\frac{1}{2}M^2+\frac{(d-4)(d-5)}{2})M^{IJ}
\end{equation}
for totally symmetric field\footnote{We use the notation
$(M^3)^{[I|J]}\equiv \frac{1}{2}M^{IK}M^{KL}M^{LJ}
-(I\leftrightarrow J)$, $M^2\equiv M^{IJ}M^{IJ}$.}.
These defining constraints can be
derived by using oscillator form of realization for $M^{IJ}$ but they
are valid, of course, for arbitrary form of realization of spin operator.

(iii) Spin operator for field of arbitrary representation, say mixed
symmetry representation, does not satisfy the constraints (\ref{defsymm}),
(\ref{defasm}). In other words the constructed generators satisfy
commutation relations of the AdS algebra provided the spin operator satisfy
the defining constraints (\ref{defsymm}), (\ref{defasm}). Therefore
the above representations for generators are valid only for totally
symmetric and antisymmetric fields.

(iv) As is well known, in the  light-cone form
the  Poincar\'e algebra generators are realized nonlinearly with respect to
$\partial^+$, namely,  $\partial^+$ appears in denominators of some
generators of Poincar\'e algebra. From the above expression it is seen
that as compared to Poincar\'e algebra generators the AdS algebra generators
take additional nonlinear dependence with respect to radial variable $z$.
Thus the AdS generators are realized non linearly in two dimensional phase
space $\partial^+$ (momentum), $z$ (coordinate).\footnote{The boost
generator $J^{+-}$ scales the momentum $p^+$ which is representation of
operator $\partial^+$ in momentum space.  In string theory in Minkowski
space-time this boost is realized as rescaling of string world sheet
coordinate $\sigma$. The world sheet scale transformations
associated with Virasoro algebra  play defining role in string
theory. In AdS space-time in addition to the scaling of $p^+$ there is the
scaling generated by dilatation generator $D$. This $D$ scales, among
others, the coordinate $z$.  One can speculate that in string theory (and
higher spin massless field theory) in AdS space-time there is an  underlying
invariance and  thus an infinite dimensional algebra
 related to this scaling.}
Note that at the same time the generators are still local in coordinate
$x^-$ and derivative with respect to $z$.

  (v) If we restore  the
dependence on cosmological constant $\lambda$ then making use of
(\ref{contract2}) one can make sure that as cosmological constant tends to
zero the light-cone generators of AdS algebra become the ones of the
Poincar\'e algebra.

The above expressions give realization of AdS algebra generators as
differential operators acting on physical fields. Now let us write down
the realization of AdS algebra generators in terms of physical fields.
As we mentioned above the kinematical generators $\hat{G}^{kin}$ are
realized quadratically in the physical fields while the dynamical
$\hat{G}^{dyn}$ are realized non-linearly. At a quadratical level both
$\hat{G}^{kin}$ and $\hat{G}^{dyn}$ have the following representation

$$
\hat{G}=\int dx^-d^{d-2}x\langle\partial^+\phi|G|\phi\rangle\,,
$$
where $G$ are the differential operators given above. The field
$|\phi\rangle$ satisfies the Poisson-Dirac commutation relation

$$
[\,|\phi(x)\rangle,\langle\phi(x^\prime)|\,]|_{equal\, x^+}
=-\frac{1}{2\partial^+}\delta(x^--x^{-\prime})
\delta^{d-3}(x-x^\prime)\delta(z,z^\prime)\,.
$$
With these definitions one has the standard commutation relation

$$
[|\phi\rangle,\hat{G}]=G|\phi\rangle\,.
$$

\newsection{General  light-cone  formalism   in AdS
space-time}

In  the previous sections we have developed light-cone formulation starting
with gauge invariant equations of motion. This strategy is
difficult to realize in many cases because the gauge invariant
formulations are not available in general. One of attractive features
of light-cone formalism is that it allows to formulate field dynamics
without knowledge of covariant formulation. Moreover,  sometimes a theory
formulated within this formalism turns out to be a good starting point for
deriving a Lorentz covariant formulation.  Derivation of covariant
formulation of string field theories is one of the
famous examples of exploiting this strategy (see \cite{SIEG,HATA}). The
practice we have got while deriving light-cone formulation for particular
cases allows us to develop general light-cone formalism  in AdS space-time.
In this section we construct light-cone form of AdS algebra generators
which are applicable to arbitrary symmetry representations, so called mixed
symmetry representations. We show that these generators can be constructed
in terms of spin operators and AdS mass operator.  We find closed defining
equations for the AdS mass operator.\footnote{A concrete representation for
this operator will be  found by applying group theoretic approach in the
next section.}

An  attractive feature
of the representation for generators we find is that they are valid
(i) for
massless and massive fields; (ii) for spin multiplets, i.e.  these
equations are in principle
applicable to string theory;
(iii) for supersymmetric
theories.
 We also demonstrate how  the light-cone description for
totally symmetric and antisymmetric fields can be derived entirely within
the light-cone formalism without using gauge invariant equations of
motion.

Generators described in the previous section have been given for arbitrary
value of evolution parameter $x^+$. As is well known,  the generators for
arbitrary $x^+$ can be obtained from the ones at  $x^+=0$ by using
the relation

$$
G_{x^+}=e^{x^+P^-}G|_{x^+=0}e^{-x^+P^-}\,.
$$  From this and from commutation relations we learn that the following
generators

$$
P^i\,,
\qquad
P^+\,,
\qquad
P^-\,,
\qquad
J^{ij}\,,
\qquad
J^{-i}\,,
\qquad
K^-
$$
do not depend on $x^+$ while for the remaining generators one has

$$
J^{+i}=J^{+i}|_{x^+=0}+x^+P^i\,,
\qquad
J^{+-}=J^{+-}|_{x^+=0}+x^+P^-\,,
\qquad
D=D|_{x^+=0}+x^+P^-\,,
$$
$$
K^i=K^i|_{x^+=0}-x^+J^{-i}\,,
\qquad
K^+=K^+|_{x^+=0}+x^+(D+J^{+-})|_{x^+=0}+x^{+2}P^-\,.
$$
Thus without loss of generality we can put $x^+=0$.  Below in this section
all the generators are considered for $x^+=0$. To develop general
light-cone we should make an assumption about form of generators. Based on
our previous study we make the following assumptions.

(i) Taking into account that the kinematical generators take the same form
for both symmetric and antisymmetric representations (with appropriate
spin operator $M^{IJ}$) we suppose that they maintain this form for
arbitrary representations

\begin{eqnarray}
\label{pig}
&&
P^i=\partial^i\,,
\\
&&
P^+=\partial^+\,,
\\
\label{jpig}
&&
J^{+i}=-x^i\partial^+\,,
\\
&&
J^{ij}=x^i\partial^j-x^j\partial^i+M^{ij}\,,
\\
\label{jmpg}
&&
J^{+-}=-x^-\partial^+\,,
\\
\label{dg}
&&
D=x^-\partial^++x^I\partial^I+\frac{d-2}{2}\,,
\\
\label{kig}
&&
K^i=-\frac{1}{2}x_J^2\partial^i+x^iD +M^{iI}x^I\,,
\\
\label{kpg}
&&
K^+=-\frac{1}{2}x_I^2\partial^+\,,
\end{eqnarray}
where the spin part $M^{IJ}$ should be taken in an  appropriate
representation.

(ii) The dynamical generator $P^-$ has the following form

\begin{equation}
\label{pmg}
P^-=-\frac{\partial_I^2}{2\partial^+}
+\frac{1}{2z^2\partial^+}A\,.
\end{equation}
Note that we do not make assumptions about the form of remaining generators
$K^-$ and $J^{-i}$. Now the problem we are going to solve here is
formulated as follows. Given spin operators $M^{IJ}$ find the AdS mass
operator $A$ and the remaining generators $J^{-i}$, $K^-$. Below
by exploiting only the commutation relations of the AdS algebra we
demonstrate that the above assumptions turn out to be sufficient to
evaluate remaining generators and get closed defining equations for AdS
mass operator $A$.

First of all from computation relations of $P^-$ with the
kinematical generators given in (\ref{pig})-(\ref{kpg}) we conclude that
the operator $A$ is independent of space-time coordinates $x^I$, $x^-$ and
their derivatives $\partial^I$, $\partial^+$, and commutes with spin
operator $M^{ij}$

\begin{equation}
\label{adef0}
[A,M^{ij}]=0\,.
\end{equation}
Second, from commutator

$$
[P^-,K^i]=-J^{-i}
$$
we find representation for $J^{-i}$

\begin{equation}
\label{jmig}
J^{-i}=x^-\partial^i -x^i P^- +M^{iJ}\frac{\partial^J}{\partial^+}
-\frac{1}{2z\partial^+}[M^{zi},A]\,.
\end{equation}
Using (\ref{pmg}) and (\ref{jmig}) we  first find

$$
[P^-,J^{-i}]=\frac{1}{4z^3\partial^{+2}}
\Bigl(-2\{M^{zi},A\}+[[M^{zi},A],A]\Bigr)\,.
$$ On  the other hand because of AdS algebra commutation relation
$[P^-,J^{-i}]=0$ we conclude that the AdS mass operator $A$ should
satisfy the following constraint

\begin{equation}
\label{adef1}
2\{M^{zi},A\}-[[M^{zi},A],A]=0\,.
\end{equation}
Making use of (\ref{kig}) and (\ref{jmig}) we evaluate then the commutator

\begin{eqnarray*}
[K^i,J^{-j}]
&=&
-\delta^{ij}(-\frac{1}{2}x_K^2P^-+x^-D
+\frac{1}{\partial^+}x^I\partial^J M^{IJ}
-\frac{x^l}{2z\partial^+}[M^{zl},A]) \\
&+&\frac{1}{2\partial^+}(\{M^{iL},M^{Lj}\}+[M^{zi},[M^{zj},A]])
\end{eqnarray*} From this and the AdS algebra commutation relation
$$
[K^i,J^{-j}]=-\delta^{ij}K^-
$$
we find the following representation for the generator $K^-$

\begin{equation}
\label{kmg}
K^-=-\frac{1}{2}x_I^2P^- +x^-D +\frac{1}{\partial^+}x^I\partial^J M^{IJ}
-\frac{x^i}{2z\partial^+}[M^{zi},A]
+\frac{1}{\partial^+}B\,,
\end{equation}
provided the spin operators $M^{IJ}$, AdS mass operator $A$ and new
operator $B$ satisfy the constraint

\begin{equation}
\label{adef2}
[M^{zi},[M^{zj},A]]+\{M^{iL},M^{Lj}\}
=-2\delta^{ij}B\,.
\end{equation}
Note that this constraint gives definition of operator $B$ in terms of
basic operators which are spin operator $M^{IJ}$ and AdS mass operator
$A$.

Thus we have derived representation for all generators of the AdS
algebra.  One can make sure that remaining commutation relations of the
algebra are also satisfied. In fact, it is sufficient to check the
commutation relation

\begin{equation}
\label{jmijmj}
[J^{-i},J^{-j}]=0\,.
\end{equation}
All others will then be satisfied because of Jacobi  identities.
Calculating the commutator in question we get

$$
[J^{-i},J^{-j}]
=\frac{1}{2z^2\partial^{+2}}
(-M^{ij}A-M^{zj}AM^{zi}+M^{zi}AM^{zj}
+\frac{1}{2}[[M^{zi},A],[M^{zj},A]])\,.
$$
Making use of Jacobi identities and defining equations
(\ref{adef1}), (\ref{adef2}) one finds

\begin{equation}\label{impcom}
[[M^{zi},A],[M^{zj},A]]
=2M^{ij}A-2M^{zi}AM^{zj}+2M^{zj}AM^{zi}\,,
\end{equation}
i.e. the commutation relation (\ref{jmijmj}) is satisfied and there are no
additional constraints on operator $A$.

Let us now summarize the results. Kinematical generators for
arbitrary representations of AdS algebra are given in
(\ref{pig})-(\ref{kpg}), while the dynamical generators are given by

\begin{eqnarray}
\label{genpm}
&&
P^-=-\frac{\partial_I^2}{2\partial^+}
+\frac{1}{2z^2\partial^+}A\,,
\\
\label{genjmi}
&&
J^{-i}=x^-\partial^i-x^iP^-+M^{iJ}\frac{\partial^J}{\partial^+}
-\frac{1}{2z\partial^+}[M^{zi},A]\,,
\\
\label{genkp}
&&
K^-=-\frac{1}{2}x_I^2P^-+x^-D +\frac{1}{\partial^+}x^I\partial^J M^{IJ}
-\frac{x^i}{2z\partial^+}[M^{zi},A]
+\frac{1}{\partial^+}B\,,
\end{eqnarray}
where AdS mass operator $A$ is \ \ \   (i)  independent of $x^I,x^-$
$\partial^I,\partial^+$, i.e. depends only on spin degree of
freedom; (ii) invariant of spin part of the $so(d-3)$ algebra
(\ref{adef0}), and  (iii) satisfies the defining equations

\begin{eqnarray}
\label{defcon1}
&&
2\{M^{zi},A\}-[[M^{zi},A],A]=0\,,
\\
\label{defcon2}
&&
[M^{zi},[M^{zj},A]]+\{M^{iL},M^{Lj}\}=-2\delta^{ij}B\,.
\end{eqnarray}
An attractive feature of the representation for
generators in (\ref{genpm})-(\ref{genkp}) as well as defining equations is
that they are  valid for massive fields, spin multiplets and supersymmetric
theories.\footnote{Obviously, supersymmetry imposes additional constraints
on the operator$A$.}

Thus the general strategy of finding light-cone form
of generators consists of the following steps:  (i) given spin, or spin
multiplet, choose appropriate spin matrix $M^{IJ}$; (ii) construct the most
general operator $A$ which commutes with $M^{ij}$; (iii) find a solution to
constraints (\ref{defcon1}), (\ref{defcon2}).

Making use of this representation we can then compute the second
order Casimir operator of AdS algebra

$$
Q=\frac{1}{2}J_{ab}^2+2K^a P^a -D(D+1-d)
$$
and find that all dependence on the space-time coordinates (orbital part)
drops, i.e. the Casimir operator is entirely determined by the spin
operator $M^{ij}$ and AdS mass operator $A$:

\begin{equation}\label{casorb}
Q=-A+2B+\frac{1}{2}M_{ij}^2+\frac{d(d-2)}{4}\,.
\end{equation}
Note that this fact is true for Poincar\'e algebra. Thus the equations of
motion take the form

\begin{equation}\label{eqmotgt}
\Bigl(-A+2B+\frac{1}{2}M_{ij}^2+\frac{d(d-2)}{4}-\langle Q\rangle
\Bigr)|\phi\rangle=0\,.
\end{equation}
where $\langle Q\rangle$ is eigenvalue of the Casimir operator $Q$ in
certain representation of the AdS algebra.

Before proceeding let to show in how manner these results can be used to
demonstrate the fact that a field in the $d$ dimensional $AdS$ space-time
irrespective of its mass can be interpreted as a field in
$(d-1)$-dimensional Minkowski space-time with continuous mass spectrum.
Toward this end let us rewrite the generators $P^-$ and $J^{-i}$ as
follows

\begin{eqnarray}
\label{pmcms}
&& P^-=-\frac{\partial_i^2}{2\partial^+}
+\frac{1}{2\partial^+}\hat{M}\,,
\\
\label{jmicms}
&&
J^{-i}=x^-\partial^i - x^iP^- +M^{ij}\frac{\partial^j}{\partial^+}
-\frac{1}{\partial^+}\hat{M}^i\,,
\end{eqnarray}
where we introduce the following operators

\begin{equation}
\label{ma}
\hat{M}\equiv -\partial_z^2+\frac{1}{z^2}A\,,
\qquad
\hat{M}^i\equiv M^{zi}\partial_z+\frac{1}{2z}[M^{zi},A]\,.
\end{equation}
The operator $\hat{M}$ in (\ref{pmcms}) can be interpreted
as mass operator for a field propagating in $(d-1)$ dimensional Minkowski
space-time while $\hat{M^i}$ is associated with spin operator of this
field.  We shall refer the $\hat{M}$ as Poincar\'e mass operator. To support
 this interpretation we should verify that these operators satisfy
appropriate commutation relations between Poincar\'e mass operator and spin
operator $\hat{M}^i$ \cite{SIEG}.  Indeed, making use of the defining
equations (\ref{defcon1}) and relationship (\ref{impcom}) one finds that
the operators $\hat{M}$ and $\hat{M}^i$ satisfy the following commutation
relations

$$
[\hat{M},\hat{M}^i]=0\,,
\qquad
[\hat{M}^i,\hat{M}^j]=M^{ij}\hat{M}\,.
$$
These commutation relations coincide with the ones for mass and spin
operators (see \cite{SIEG}). Therefore the generators $P^-$, $J^{-i}$
given in (\ref{pmcms}), (\ref{jmicms}) coincide exactly with ones for
massive field in $(d-1)$-dimensional Minkowski space-time.
Thus we have demonstrated that field in the $d$-dimensional AdS
space-time can indeed be interpreted as massive field in $(d-1)$ dimensional
space-time.\footnote{This fact was discussed in literature only for
scalar field.  Our formalism allows us to demonstrate this fact explicitly
for arbitrary spin fields at the level of generators matching
(\ref{ma}).}  In this matching the AdS algebra becomes the
algebra of conformal transformation of boundary Minkowski space-time. The
fact that in $(d-1)$ dimensions the spectrum is continuous follows from
the relation

$$
e^{x D}P_a^2 e^{-x D}= e^{-2x} P_a^2
$$
(see \cite{macksal}). Interesting point of our study is that it allows us
to demonstrate explicitly how do the Poincar\'e mass and spin
operators of massive field in Minkowski space-time relate with AdS mass and
spin operator of field in AdS space-time
(\ref{ma})\footnote{As a side a remark let us note that
this fact might have interesting application to a problem of interrelation
of higher spin massless spin fields living in $d=11$ - dimensional AdS
space-time, which is a bulk, and superstring theories living at $d=10$
Minkowski space-time, which is a boundary.  One can conjecture that
string theory can be interpreted as resulting from some kind of a
spontaneous breakdown of symmetries of higher spin massless fields theory.
The problem of continuous mass spectrum could be solved then by
spontaneous breakdown of original global AdS symmetries.}.

As a illustration how does the general light-cone formalism work
let us rederive the representation of AdS algebra for symmetric and
antisymmetric fields. Note that in previous section we derived this
representation by using gauge invariant equations of motion. Now we start
directly with light-cone formalism and look for solution to defining
equations (\ref{defcon1}), (\ref{defcon2}).
Toward this end we look for the following form of operator $A$:

$$
A=aM_{ij}^2+b\,,
$$
where $a$ and $b$ arbitrary functions of spin Casimir operator $M_{IJ}^2$.
We are going to find these $a$ and $b$  by applying the defining equations.
The above operator $A$ obviously commutes with $M^{ij}$. Next step is to
analyse the defining constraint (\ref{defcon1}). This constraint takes the
following form

\begin{eqnarray}
\label{defequapl}
&&
2\{M^{zi},A\}-[[M^{zi},A],A]
\\
&&
=-16a^2(M^3)^{[z|i]}+2a(1+2a)\{M^{zi},M_{kl}^2\}
+(-8a^2 M^2+4a^2(N-2)^2+4b)M^{zi}=0\,.
\nonumber
\end{eqnarray}
At this point we should use defining equations for spin operators $M^{IJ}$.
By using the defining equations for spin operator of totally antisymmetric
field (\ref{defasm}) we express the first term in r.h.s. of equation
(\ref{defequapl}) in terms of others. From this we get solution

\begin{equation}
\label{soldefas}
a=-\frac{1}{2}\,,
\qquad
b=M^2+\frac{d(d-2)}{4}\,,
\end{equation}
Exploiting defining equations for spin operator of symmetric
field (\ref{defsymm}) in equation (\ref{defequapl}) leads to

\begin{equation}
\label{soldefsym}
a=-\frac{1}{2}\,,
\qquad
b=\frac{(d-4)(d-6)}{4}\,.
\end{equation}  From the second defining equations (\ref{defcon2})
we get the equations

$$
4a\delta^{ij}M_{zl}^2+(1+2a)\{M^{iL},M^{Lj}\}
=-2\delta^{ij}B
$$
and taking into account (\ref{soldefas}) and (\ref{soldefsym})
we get

$$
B=M_{zi}^2
$$
for both cases.  These results coincide exactly with the those of
the previous
sections. Here we derived them in purely algebraic way without  using
covariant equations of motion. This derivation is much  simpler and
demonstrates efficiency of light-cone approach.
Note that in this derivation we used only the defining
equations for spin matrices $M^{IJ}$ (\ref{defsymm}),(\ref{defasm}).
This is all what is required to get a
light-cone description of a specific fields  from general
light-cone formalism.

\newsection{${}~\!\!\!$Light-cone generators for arbitrary representations
of AdS algebra. Group theoretic approach}

In this section we develop light-cone form of AdS generators
both for massless and massive fields. To do that we use method of induced
representations (for reviews see \cite{mackey},\cite{mensky}). Let us
briefly describe this method. Let $g$ be a group and $h$ its subgroup.
The decomposition $g=hg_x$ defines $g_x$ as coset representative of $g/h$.
In the following  the generators of $h$ and $g_x$ are denoted by ${\bf H}$
and ${\bf K}$ respectively. To define induced representation one introduces
a space of functions $\varphi$ subject to the following condition

\begin{equation}\label{strcon}
\varphi(hg_x)=\Delta(h)\varphi(g_x)\,,
\end{equation}
where $\Delta$ is a representation of subgroup $h$

\begin{equation}\label{hmrr}
\Delta(h_1h_2)=\Delta(h_1)\Delta(h_2)\,.
\end{equation}
Now the induced representation $T_g$ is defined by

\begin{equation}\label{tconr0}
(T_{g_1}\varphi)(g)=\varphi(gg_1)\,.
\end{equation}
The condition (\ref{strcon}) tells that the functions $\varphi$ are
defined on coset space. It is convenient to introduce an unconstrained
function $\phi$ defined by $\phi(x)=\varphi(g_x)$. In terms of
$\phi(x)$ the formula (\ref{tconr0}) takes then the form

\begin{equation}\label{tconr}
(T_{g_1}\phi)(x)=\Delta(h(x,g_1))\phi(xg_1)\,,
\end{equation}
where $h(x,g)$ and $xg$ are defined from a decomposition

\begin{equation}\label{colar}
g_xg=h(x,g)g_{xg}\,.
\end{equation}
Since we are interested in differential form for generators we consider
the formula (\ref{tconr}) for infinitesimal transformations

$$
g_1=1+\epsilon_M G^M\,,
\qquad
\epsilon \ll 1\,.
$$
We use indices $M,N,P$ to label all generators of algebra $G^M$, while the
indices $\alpha,\beta,\gamma$ label generators of subalgebra
${\bf H}$. For infinitesimal transformations we have

$$
(xg_1)^{\hat{\mu}}=x^{\hat{\mu}}+\epsilon_M\xi^{M \hat{\mu}}\,,
\qquad
h(x,g)=1+\epsilon_M \Omega^M_\alpha H^\alpha\,,
$$
where $\xi^{M\hat{\mu}}$ and $\Omega^M_\alpha$ are Killing vectors
and $h$-compensators respectively. They satisfy the relations

\begin{eqnarray}
\label{keqr}
[(\xi^M\partial),(\xi^N\partial)]=f^{MN}{}_P(\xi^P\partial)\,,
\quad
(\xi^M\partial)\Omega^N_\gamma-(\xi^N\partial)\Omega^M_\gamma
=f^{MN}{}_P\Omega^P_\gamma
-f^{\alpha\beta}{}_\gamma \Omega^M_\alpha\Omega^N_\beta\,,
\end{eqnarray}
where
$(\xi^M\partial)\equiv\xi^{M\hat{\mu}}\partial_{x^{\hat{\mu}}}$.
The $x^{\hat{\mu}}$ are coordinates of coset space $g_x$. These
coordinates should not be confused with the coordinates $x^\mu$ used
in previous sections to describe AdS space-time.
Useful formula for calculation of Killing vectors and
$h$-compensators is

\begin{equation}\label{kvcor}
g_xG^Mg_x^{-1}=\xi^{M\hat{\mu}} \partial_{\hat{\mu}} g_x g_x^{-1}
+\Omega^M_\alpha H^\alpha\,.
\end{equation}
Using these relations and taking into account
$T_{g_1}=1+\epsilon_M G^M$
we get from (\ref{tconr}) the following representation for generators $G^M$
in terms differential operators acting on $\phi(x)$

\begin{equation}\label{difope}
G^M=\xi^{M \hat{\mu}} \partial_{\hat{\mu}}
+\Omega^M_\alpha \Delta^\alpha\,,
\end{equation}
where spin operators $\Delta^\alpha$ defined by
$\Delta^\alpha\equiv \partial^\alpha \Delta(h)|_{h=1}$
satisfy the commutation relations

\begin{equation}
\label{defi}
[\Delta^\alpha,\Delta^\beta]=f^{\alpha\beta}{}_\gamma\Delta^\gamma\,.
\end{equation}
Using (\ref{keqr}) it is easy  to
demonstrate that the generators $G^M$ given in (\ref{difope})
satisfy the commutation relations

$$
[G^M,G^N]=f^{MN}{}_PG^P\,.
$$
In next section we use the basic formulas (\ref{kvcor}) and (\ref{difope})
to obtain group theoretic representation of AdS algebra generators.

\subsection{ Light-cone form of generators of AdS algebra. Group theoretic
representation}

To apply the  method of induced representations we should fix
decomposition of AdS algebra into subalgebra ${\bf H}$ and coset space
algebra ${\bf K}$.  To do that we use Iwasawa decomposition
${\bf G}={\bf H}^m{\bf AN}$, where ${\bf H}^m$ is a maximal compact algebra,
${\bf A}$ is a maximal abelian algebra in ${\bf G}/{\bf H}^m$ and
${\bf N}$ nilpotent algebra in ${\bf G}/{\bf H}^m$. The ${\bf N}$
transforms in certain representation of ${\bf A}$.  Next,  ${\bf G}$
can be decomposed as ${\bf G}={\bf K}{\bf ANC}$, where ${\bf C}$ are
elements of ${\bf H}^m$ that commute with ${\bf A}$, while  ${\bf K}$
is the  remainder of ${\bf H}^m$: ${\bf K}={\bf H}^m/{\bf C}$.  For the
case of AdS algebra we have

\begin{eqnarray}
\label{nilpot}
&
{\bf A}=D,\,J^{+-}\,,
\qquad
{\bf N}=P^-,\,J^{-i},\,K^-,\,K^i\,,
\qquad
{\bf C}=J^{ij}\,,
&
\\
\label{cosgen}
&{\bf K}= P^i,\, P^+,\, J^{+i},\, K^+.&
\end{eqnarray}
The subalgebra ${\bf H}$ is then ${\bf G}/{\bf K}$, i.e.
${\bf H}={\bf ANC}$ and in terms of generators it looks as
\begin{equation}
\label{subgen}
{\bf H}=P^-,\,J^{-i},\,K^-,\,K^i,\, D,\, J^{+-},\, J^{ij}\,.
\end{equation}
For AdS algebra the subalgebra ${\bf H}$ is analog of little group.
Note that subalgebra ${\bf H}$ and coset space generators ${\bf K}$ can be
represented as follows
$$
{\bf K}= (P^+,J^{+i},K^+)
{\,\lower-0.2ex\hbox{${\scriptstyle+}$}}
{\kern-1.3ex\hbox{$\supset$}\,}
P^i\,,
\qquad
{\bf H}=\Bigl((P^-,J^{-i},K^-)
{\,\lower-0.2ex\hbox{${\scriptstyle+}$}}
{\kern-1.3ex\hbox{$\supset$}\,}
K^i\Bigr)
{\,\lower-0.2ex\hbox{${\scriptstyle+}$}}
{\kern-1.3ex\hbox{$\supset$}\,}
(D,J^{+-},J^{ij})\,.
$$
Thus the coset space associated with ${\bf K}$ is used as carrier of
representation, while ${\bf H}$ is used as inducing subgroup.

The coset representative $g_x$ and coset coordinates we use are given by
\begin{equation}
\label{ggogt}
g_x=g_1g_2\,,
\qquad\qquad
g_1\equiv \exp(x^-P^+ +v^iJ^{+i}+z^-K^+)\,,
\qquad
g_2\equiv \exp(x^iP^i)\,.
\end{equation}
The wavefunctions will depend on coordinates $x^i,v^i,x^-,z^-$, which we
refer as group theoretic coordinates. After certain
redefinitions the coordinates $x^i$ and $x^-$ will be related with similar
coordinates of field theoretical approach, while the coordinate $z^-$ will
be related with $z$.

First we calculate the coset space generators  (\ref{cosgen}) by applying
the basic formula (\ref{difope}).
For illustration purposes let us present the calculation of coset
generators in details.  To apply the formula (\ref{difope}) we need
Killing vectors and $h$-compensators which can be obtained from
(\ref{kvcor}). To this end we start with the following relation for right
invariant form

$$
dg_xg_x^{-1}=(dv^i-z^-dx^i)J^{+i}+(dx^-+v^idx^i)P^+ + dz^- K^+ +P^idx^i\,.
$$ From this we get

$$
\partial_{x^-}g_xg_x^{-1}=P^+\,,
\quad
\partial_{v^i}g_xg_x^{-1}=J^{+i}\,,
\quad
\partial_{z^-}g_xg_x^{-1}=K^+\,,
\quad
\partial_{x^i}g_xg_x^{-1}=P^i +v^iP^+ -z^-J^{+i}\,.
$$
Using then the relations
\begin{eqnarray*}
&&
g_xP^+g_x^{-1}=P^+\,,
\hspace{5cm}
g_xJ^{+i}g_x^{-1}=J^{+i}-x^i P^+\,,
\\
&&
g_xK^+g_x^{-1}=K^+ -\frac{1}{2}x_j^2P^+ +x^iJ^{+i}\,,
\qquad\qquad
g_xP^ig_x^{-1}=P^i +v^iP^+ - z^-J^{+i}\,,
\end{eqnarray*}
and (\ref{kvcor}) we get the Killing vectors
$$
\xi^{P^i,x^j}=\delta^{ij}\,,
\quad
\xi^{P^+,x^-}=1\,,
\quad
\xi^{J^{+i},v^j}=\delta^{ij}\,,
\quad
\xi^{J^{+i},x^-}=-x^i\,,
$$
$$
\xi^{K^+,v^i}=x^i\,,
\quad
\xi^{K^+,x^-}=-\frac{1}{2}x_j^2\,,
\qquad
\xi^{K^+,z^-}=1\,,
$$
while corresponding $h$-compensators are equal to zero.
Note that these Killing vectors should not be confused with the ones
in (\ref{kilvec}) which describe AdS space-time transformations. Making use
of these expressions and formula (\ref{difope}) we get the following
representation for coset generators ${\bf K}$:

\begin{eqnarray}
\label{pigt}
&&
P^i=\partial_{x^i}\,,
\\
&&
P^+=\partial_{x^-}\,,
\\
\label{jpigt}
&&
J^{+i}=-x^i\partial_{x^-} + \partial_{v^i}\,,
\\
\label{kpgt}
&&
K^+=-\frac{1}{2}x_j^2\partial_{x^-}
+x^i\partial_{v^i}+\partial_{z^-}\,.
\end{eqnarray}
Now let us derive the representation for generators of
the subalgebra ${\bf H}$ (\ref{subgen}).
In accordance with general prescription of formula
(\ref{difope}) we should introduce spin operators $\Delta^\alpha$ for each
generator of subalgebra ${\bf H}$ (\ref{subgen})

$$
\Delta^{P^-},\,
\qquad
\Delta^{J^{-i}},\,
\qquad
\Delta^{K^-},\,
\qquad
\Delta^{K^i},\,
\qquad
\Delta^D,\,
\qquad
\Delta^{J^{+-}},\,
\qquad
\Delta^{J^{ij}}\,.
$$
These operators satisfy, by definition, the same commutation relations as
generators of subalgebra ${\bf H}$ (\ref{subgen}).
Making use of the basic formula
(\ref{difope}) we get then the following representation for the generators
of the subalgebra ${\bf H}$

\begin{eqnarray}
\label{jpmgt}
&&
{}~\hspace{-1cm}
J^{+-}=-v^i\partial_{v^i}-x^-\partial_{x^-}-z^-\partial_{z^-}
+\Delta^{J^{+-}}\,,
\\
\label{dgt}
&&
{}~\hspace{-1cm}
D=x^i\partial_{x^i}+x^-\partial_{x^-}-z^-\partial_{z^-}
+\Delta^D\,,
\\
&&
{}~\hspace{-1cm}
J^{ij}=x^i\partial_{x^j}-x^j\partial_{x^i}
+v^i\partial_{v^j}-v^j\partial_{v^i}+\Delta^{J^{ij}}\,,
\\
\label{kigt}
&&
{}~\hspace{-1cm}
K^i=-\frac{1}{2}x_j^2\partial_{x^i}
+x^i D
+(v^i\partial_{v^j}-v^j\partial_{v^i}+\Delta^{J^{ij}})x^j
-x^-\partial_{v^i}+v^i\partial_{z^-}
+\Delta^{K^i}\,,
\\
\label{pmgt}
&&
{}~\hspace{-1cm}
P^-=-v^i\partial_{x^i}
+\frac{1}{2}v^2\partial_{x^-}+(z^-)^2\partial_{z^-}
-(\Delta^D+\Delta^{J^{+-}})z^-+\Delta^{P^-}\,.
\end{eqnarray}
Note that explicit expressions for the remaining generators
$K^-$ and $J^{-i}$ can be found from above expressions for
$P^-$, $K^i$ and commutation relations of the $so(d-1,2)$ algebra.
The expressions for the generators given in (\ref{pigt})-(\ref{kpgt}) and
(\ref{jpmgt})-(\ref{pmgt}) provide group theoretic representation for
light-cone form of the AdS algebra generators.

\subsection{Interrelation between group theoretic and field theoretic
approaches. Solution to defining equations}

Group theoretic representation for generators is constructed in a
basis
which differs from that used in field theoretic approach. Group theoretic
representation is given in terms of coset space coordinates -- group
theoretic coordinates -- defined by the relations (\ref{ggogt}), while in
the field theoretic approach we used the  usual space-time coordinates.
Therefore in order
to match these different forms we should transform them to the same basis.

We shall transform group theoretic representation to the one of
field theoretic.
There is a number of reasons for that. One reason is
that we have formulated defining equations for generators written
in
space-time coordinates. Therefore to find the AdS mass operator and spin
operator it is necessary match orbital parts of generators. Another reason
why we consider the field theoretic formulation as the
preferable one, which
is behind this work, is related to interaction vertices. To find
interaction vertices one imposes locality condition, which implies that
the vertices should be polynomial in derivatives with respect to
transverse space-time coordinates. In other words,  formulation in terms
of space-time coordinates is preferable because in this form one can apply
locality condition directly. So our aim here is to find transformation
which match orbital part of group theoretic generators and the field
theoretic ones. After that by comparing spin part we find immediately
representation for AdS mass operator $A$ as well as spin operators
$M^{IJ}$. This representation gives solution to defining equations of
the previous section.

Before proceeding let us make some simplifications. We are interested
in irreducible representation of AdS algebra. According to the general
theory (see \cite{mackey},\cite{mensky}) in order to get such a
representation it is sufficient to set spin operators corresponding the
nilpotent algebra ${\bf N}$ (\ref{nilpot}) equal to zero

\begin{equation}\label{zspin}
\Delta^{P^-}=\Delta^{J^{-i}}=\Delta^{K^-}=\Delta^{K^i}=0\,.
\end{equation}
For remaining spin operators we use the notation

\begin{equation}\label{nzspin}
\Delta^{J^{+-}}=-\lambda^{+-}\,,
\qquad
\Delta^D=-\lambda\,,
\qquad
\Delta^{J^{ij}}=m^{ij}\,,
\end{equation}
where the $\lambda^{+-}$ and $\lambda$ are $c$-numbers, while spin operator
$m^{ij}$ satisfies commutation relations of $so(d-3)$ algebra

$$
[m^{ij},m^{kl}]=\delta^{jk}m^{il}+3 \hbox{ terms}\,.
$$
$\lambda$ in (\ref{nzspin}) and below should not be confused with
cosmological constant.
Thus we shall use the generators (\ref{jpmgt})-(\ref{pmgt}) with spin
operators given in (\ref{zspin}), (\ref{nzspin}).
Before  proceeding  it is important to understand which group theoretic
variable is responsible for spin degrees of freedom. Below we show that
the spin degrees of freedom are described by the variable $v^i$
together with spin operator $m^{ij}$.  To be precise,  let us make the
following Fourier transform

$$
f(x^i,x^-,z^-,v^i)=\int du \exp(-uv\partial^+)f_1(x^i,x^-,z^-,u^i)\,,
$$
where the $f_1$ is a new wavefunction. In basis of $f_1$
the coordinate $v^i$ takes the representation

$$
v^i \rightarrow \frac{\partial_{u^i}}{\partial^+}\,,
\qquad
\partial_{v^i}\rightarrow -u^i\partial^+\,.
$$
It turns out that in group theoretic approach it is the new variable $u^i$
that is an analog of oscillators used in field theoretic approach to
describe spin degrees of freedom. Remaining group theoretic
variables $x^i$, $x^-$, $z^-$ are related to space-time coordinates of field
theoretic approach. Let us now describe this interrelation in  more detail.

First,  we are trying to match coset generators (\ref{cosgen})
given in (\ref{pig})-(\ref{kpg}) and (\ref{pigt})-(\ref{kpgt}). It is
obvious that the generators $P^i$ and $P^+$ already coincide. Next step
is to match the generators $J^{+i}$.  To do that we choose instead of
the basis $f_1$ the following new $f_2$ basis defined by

$$
f_1(x^i,x^-,z^-,u^i)=e^{u\partial_x}
f_2(x^i,x^-,z^-,u^i)\,.
$$
Taking into account that in this new basis the coordinates take
the representation

$$
x^i\rightarrow x^i-u^i\,,
\qquad
v^i\rightarrow \frac{1}{\partial^+}
(\partial_{u^i}+\partial_{x^i})\,,
\qquad
\partial_{v^i}\rightarrow -u^i\partial^+
$$
one can make sure that the generator $J^{+i}$ (\ref{jpigt}) takes the
desired form given in (\ref{jpig}). After that we should match the group
theoretic $K^+$ (\ref{kpgt}) and the one of field theoretic. To do that we
choose instead of $f_2$ the following new $f_3$ basis

$$
f_2(x^i,x^-,z^-,u^i)=
\int dz \, z\exp(-\frac{1}{2}z^-\partial^+(z^2+u^2))
f_3(x^i,x^-,z,u^i)\,.
$$
In this new basis the generator $J^{+i}$ is not changed, while the
generator $K^+$ takes the desired form of field theoretic approach given
in (\ref{kpg}). Thus at this stage group theoretic coset generators match
with the ones of field theoretic.

Second step is to analyse the group theoretic dilatation generator
$D$ (\ref{dgt}) which in basis of $f_3$ takes the form

$$
D=x\partial_x+x^-\partial^++z\partial_z+u\partial_u+ d-1-\lambda\,.
$$
This $D$ consists of unwanted term $u\partial_u$ which is absent in $D$
of field theoretic approach. To remove this term we choose the basis
$f_4$ which is related to the previous basis $f_3$ as follows

$$
f_3(x^i,x^-,z,u^i)= f_4(x^i,x^-,z,\zeta^i)\,,
\qquad
u^i\equiv\frac{1}{z}\zeta^i\,.
$$
It is straightforward to see that in the basis of $f_4$ the
unwanted term $u\partial_u$ in $D$ is cancelled and we get

\begin{equation}
\label{d4}
D=x^-\partial^++x^I\partial^I+d-1-\lambda\,.
\end{equation}
By now the orbital part of this operator coincides with the one of
field theoretic approach (\ref{dg}). Before we match spin part of $D$ let
us write down the expressions for the group theoretic generators $J^{+-}$
(\ref{jpmgt}) and $P^-$ (\ref{pmgt}) in $f_4$ basis

\begin{equation}
\label{jpm4}
J^{+-}=-x^-\partial^+ -\lambda^{+-}\,,
\qquad
P^-=-\frac{\partial_I^2}{2\partial^+}
-\frac{c^\prime+1}{2z\partial^+}\partial_z
+\frac{1}{2z^2\partial^+}
\Bigr(\partial_\zeta^2+(\zeta\partial_\zeta)^2
+c^\prime\zeta\partial_\zeta\Bigl)\,,
\end{equation}
where
\begin{equation}\label{cpr}
c^\prime\equiv d-1-2\lambda-2\lambda^{+-}\,.
\end{equation}
All that now remains is to find the basis in which the generators
$J^{+-}$, $D$, $P^-$ given in (\ref{d4}), (\ref{jpm4})
take the form of the ones given in (\ref{jmpg}), (\ref{dg})
and (\ref{pmg}). This can be done by using the following
$f_5$ basis

$$
f_4(x^i,x^-,z,\zeta^i)=
(\partial^+)^{\lambda^{+-}}
z^{-(c^\prime+1)/2}f_5(x^i,x^-,z,\zeta^i)\,.
$$
In $f_5$ basis the generators $J^{+-}$, $D$, $P^-$ take precisely the
desired form of the field theoretic approach given in (\ref{jmpg}),
(\ref{dg}), (\ref{pmg}) with the following representation for AdS mass
operator $A$

\begin{equation}\label{agt}
A=\partial_\zeta^2+(\zeta\partial_\zeta)^2+c^\prime\zeta\partial_\zeta
+\frac{c^{\prime 2}-1}{4}\,.
\end{equation}
In order to complete our analysis we should evaluate the group
theoretic generator $K^i$ (\ref{kigt}) in the basis of $f_5$. The
relatively straightforward calculations gives the desired form of field
theoretic approach given in (\ref{kig}) with the following representation
for spin operators

\begin{equation}
\label{mijgt}
M^{ij}
=\zeta^i\partial_{\zeta^j}-\zeta^j\partial_{\zeta^i} +m^{ij}\,,
\qquad
M^{zi} = \frac{1}{2}(1-\zeta^2)\partial_{\zeta^i}
+\zeta^i(\zeta\partial_\zeta-\lambda)
+m^{ij}\zeta^j\,.
\end{equation}
Note that in section 4 we have made assumption about form of
kinematical generators (\ref{pig})-(\ref{kpg}) and dynamical generator
$P^-$ (\ref{pmg}). The derivation of this section demonstrates that this
conjectured form is in fact a most general form. One can verify
that the above expressions for AdS mass operator (\ref{agt}) and spin
operator $M^{IJ}$ (\ref{mijgt})  satisfy
the defining equations (\ref{defcon1}), (\ref{defcon2}) and give the
following representation for the operator $B$

\begin{equation}
\label{bgt}
B=\frac{1}{2}(1-\zeta^2)\partial_\zeta^2
+(\zeta\partial_\zeta)^2
+\frac{c^\prime+d-5}{2}\zeta\partial_\zeta
-m^{ij}\zeta^i\partial_{\zeta^j}
-\lambda(\lambda+\frac{c^\prime+d-3}{2})\,.
\end{equation}
Thus we have transformed the group theoretic form to the basis where the
wavefunction depends on the space-time variables $x^-$, $x^I$ and spin
variables $\zeta^i$. In addition wavefunction transforms in representation
of little spin operator $m^{ij}$. In order to understand what
kind of form of realization of spin degrees of freedom we have obtained we
should inspect the spin operator $M^{IJ}$. It is straightforward to see
that the representation we derived for $M^{IJ}$ is nothing but the
stereographic form of realization of spin degrees of freedom.  This fact
is demonstrated in Appendix B. In other words,  group theoretic approach
naturally leads to the stereographic form of realization of spin degrees
of freedom. At the same time, it turns out that the above
representation for operators $A$, $B$ and $M^{IJ}$ can be put into the form
which does not rely on stereographic form of realization of spin degrees
of freedom.  To this end we introduce the following operators

\begin{eqnarray}
\label{gtsppi}
&&
{\sf p}^i=\partial_{\zeta^i}\,,
\hspace{4cm}
{\sf k}^i=-\frac{1}{2}\zeta^2\partial_{\zeta^i}
+\zeta^i(\zeta\partial_\zeta-\lambda)
+m^{ij}\zeta^j\,,
\\
&&
{\sf m}^{ij}=\zeta^i\partial_{\zeta^j}
-\zeta^j\partial_{\zeta^i}+m^{ij}\,,
\label{gtspd}
\qquad
{\sf d}=\zeta\partial_\zeta-\lambda\,,
\end{eqnarray}
These operators satisfy commutation relations of $so(d-2,1)$ algebra

$$
[{\sf d},{\sf p}^i]=-{\sf p}^i\,,
\qquad
[{\sf d},{\sf k}^i]={\sf k}^i\,,
\qquad
[{\sf p}^i,{\sf p}^j]=0\,,
\qquad
[{\sf k}^i,{\sf k}^j]=0\,,
$$

$$
[{\sf p}^i,{\sf m}^{jk}]=\delta^{ij}{\sf p}^k
-\delta^{ik}{\sf p}^j\,,
\qquad
[{\sf k}^i,{\sf m}^{jk}]=\delta^{ij}{\sf k}^k
-\delta^{ik}{\sf k}^j\,,
$$

$$
[{\sf p}^i,{\sf k}^j]=\delta^{ij}{\sf d}-{\sf m}^{ij}\,,
\qquad
[{\sf m}^{ij},{\sf m}^{kl}]
=\delta^{jk}{\sf m}^{il}+3\hbox{ terms}
$$
The remarkable fact is that the operators $A$ (\ref{agt}), $B$
(\ref{bgt}) and the spin operator $M^{IJ}$ (\ref{mijgt})
are expressible entirely in terms of above operators

\begin{equation}
\label{gtainv}
A={\sf p}^2+{\sf d}^2+(c^\prime+2\lambda){\sf d}
+\frac{(c^\prime+2\lambda)^2-1}{4}\,,
\qquad
B=\frac{1}{2}{\sf p}^2+{\sf k}{\sf p}
+(\lambda+\frac{c^\prime+d-3}{2}){\sf d}\,,
\end{equation}
\begin{equation}\label{gtmijinv}
M^{ij}={\sf m}^{ij}\,,
\qquad
M^{zi}=\frac{1}{2}{\sf p}^i+{\sf k}^i\,.
\end{equation}
To summarize we have found two realization for spin operator $M^{IJ}$ and
AdS mass operator. First realization given in (\ref{agt}),
(\ref{mijgt}) is constructed in terms stereographic coordinates. The
second realization given in (\ref{gtainv}), (\ref{gtmijinv}) is
constructed in terms of generators of $so(d-2,1)$ and does not related to
specific coordinates.

By inserting above representations for $A$ and $B$ into (\ref{casorb})
we get for the Casimir operator

\begin{equation}\label{caseig3}
-A+2B+\frac{1}{2}M_{ij}^2+\frac{d(d-2)}{4}
=-\lambda^{+-}(\lambda^{+-}+1-d)-\lambda(\lambda+d-3)
+\frac{1}{2}m_{ij}^2\,.
\end{equation}
The positive energy lowest weight irreducible representations of
$so(d-1,2)$ algebra denoted as $D(E_0, \lambda,{\bf h})$, are defined by
$E_0$, an lowest eigenvalue of the energy operator, and by
$(\lambda,{\bf h})$\footnote{
The $\lambda$ and ${\bf h}=(h_1,\ldots,h_{[(d-3)/2)]})$
as weights of the $so(d-1)$ algebra representation satisfy the
relation $\lambda \geq h_1\ldots\geq h_{[(d-3)/2]}\geq 0$ for even $d$ and
the relation $\lambda \geq h_1\ldots\geq h_{[(d-5)/2]}\geq
|h_{[(d-3)/2]}|$ for odd $d$.}
which is the weight of the $so(d-1)$ algebra representation in
$so(2)\oplus so(d-3)$ basis.  In $D(E_0,\lambda,{\bf h})$ the Casimir
operator takes the value

\begin{equation}\label{caseig4}
\langle Q\rangle=-E_0(E_0+1-d)-\lambda(\lambda+d-3)
+\frac{1}{2}\langle m_{ij}^2\rangle\,,
\end{equation}
where $\langle m_{ij}^2/2\rangle$ is a eigenvalue Casimir operator of
$so(d-3)$ algebra representation. Inserting relations (\ref{caseig3})
and (\ref{caseig4}) into (\ref{eqmotgt}) suggests the following
identification

\begin{equation}\label{ez}
E_0=\lambda^{+-}\,.
\end{equation}
Below we demonstrate that the $E_0$ defined by this relation is indeed
lowest energy value. The representation we obtained describes massive
field in general.  The basis for the spin states of massive field can be
constructed as follows.  Introduce the vector $|{\bf h}\rangle$ which is
(i) the eigenvalue vector of operator ${\sf d}$, and (ii) a weight
${\bf h}$ representation of the $so(d-3)$ algebra. This vector satisfies
the following conditions

$$
{\sf p}^i|{\bf h}\rangle=0\,,
\qquad
{\sf d}|{\bf h}\rangle=-\lambda|{\bf h}\rangle\,.
$$
The first constraint is imposed to get  a  finite dimensional
representation. Note that in representation given in
(\ref{gtsppi})-(\ref{gtspd}) these constraints imply that $|{\bf h}\rangle$
does not depend on $\zeta^i$. Next construct the space

\begin{equation}\label{massta}
\Lambda\equiv \sum_{\sigma=0}^{2\lambda} \oplus
\Lambda^\sigma\,,
\qquad
\Lambda^\sigma\equiv
{\sf k}^{i_1}\ldots {\sf k}^{i_\sigma}|{\bf h}\rangle\,,
\end{equation}
where $\Lambda^{2\lambda+1}=0$. Dimension of the space $\Lambda$
coincides with a dimension of the $so(d-1)$ algebra irreducible
representation which is used to describe spin degrees of freedom of massive
field. Therefore the $\Lambda$ is appropriate to describe
spin degrees of freedom of massive field.  Obviously the space $\Lambda$ is
invariant under the action of spin operators $M^{IJ}$ and AdS mass
operator $A$ (see (\ref{gtainv}), (\ref{gtmijinv})).  However for certain
values of $c^\prime$ there is invariant subspace in $\Lambda$.  This
invariant subspace describes spin degrees of freedom of massless field.
In order to understand this fact better it is instructive to consider
examples.

{\it Spin one Maxwell field}. Let us start with the simplests case of spin
one Maxwell field. In this case we have $\lambda=1$ and ${\bf h}=0$
\footnote{In this case the vector $|{\bf h}\rangle$
in (\ref{massta}) is simply constant which we set equal to unity and we
use the $so(d-3)$ algebra spin operator $m^{ij}=0$. Then from
(\ref{massta}) we get that $\Lambda=(1,\zeta^i,\zeta^2)$.  Dimension of
this $\Lambda$ is equal to $d-1$ which is dimension of massive spin field
$|\phi\rangle=\phi_1+\zeta^i\phi^i+\zeta^2\phi_2$. The invariant
shortened subspace appropriate for massless case is then given by
(\ref{phivec}).}. Wavefunction of spin massless field in stereographic
coordinates is given by

\begin{equation}\label{phivec}
|\phi\rangle=|\phi_1\rangle+|\phi_0\rangle\,,
\qquad
|\phi_1\rangle\equiv \zeta^i\phi^i\,,
\quad
|\phi_0\rangle\equiv (1-\zeta^2)\phi\,.
\end{equation}
The $|\phi_1\rangle$ and $|\phi_0\rangle$  transform into one another
under the action of spin operator $M^{IJ}$, i.e. the vector $|\phi\rangle$
transforms in representation of $M^{IJ}$.  This vector has $d-2$ degrees
of freedom and therefore is appropriate to describe massless spin one
field. For arbitrary $c^\prime$ the AdS mass operator $A$ moves the
$|\phi\rangle$ out the form given in (\ref{phivec}). Indeed
straightforward calculation gives

$$
A|\phi\rangle=\kappa_1|\phi_1\rangle+\kappa_2|\phi_0\rangle
-2(d-1+c^\prime)|V\rangle\,,
\qquad
|V\rangle\equiv\phi\,,
$$
where $\kappa_1$ and $\kappa_2$ are certain constants. Obviously the vector
$|V\rangle$ does not belong to the invariant subspace of $|\phi\rangle$
given in (\ref{phivec}).  The contribution of $|V\rangle$ can be cancelled
whenever $c^\prime=1-d$.  Taking into account (\ref{cpr}) and (\ref{ez}) we
find $E_0=d-2$ and this nothing but the lowest energy values for spin one
massless field in $d$-dimensional AdS space-time (\cite{metsit1}).  Thus the
value $c^\prime$ is fixed from the requirement that the $|\phi\rangle$
transforms into itself under the action of AdS mass operator $A$.

{\it Totally antisymmetric spin $s$ field}.
In this case we have $\lambda=1$ and the weights of vector
$|{\bf h}\rangle$ are given by ${\bf h}=(1,\ldots,1,0\ldots,0)$ (where the
unity occurs $s-1$ times in this sequence). The invariant subspace in
$\Lambda$ appropriate to the description of massless field is given
by the wavefunction\footnote{ The vector $|{\bf h}\rangle$, which is used
to construct the space $\Lambda$, is given by
$\alpha^{i_1}\ldots\alpha^{i_{s-1}}|0\rangle$. Then the space $\Lambda$ is
 described by expression (\ref{massta}) and is given by
$\Lambda
=({\sf k}^i{\sf k}^j|{\bf h}\rangle,{\sf k}^i|{\bf h}\rangle,
|{\bf h}\rangle)$. The ${\sf k}^i$ is given in (\ref{gtsppi})
where we should exploit the following spin operator
$m^{ij}=\alpha^i\bar{\alpha}^j-\alpha^j\bar{\alpha}^i$.}

\begin{equation}\label{pp1p0}
|\phi\rangle=|\phi_1\rangle+|\phi_0\rangle\,,
\end{equation}
where
\begin{eqnarray}
\label{phi1as}
&&
{}~\hspace{-1.5cm}
|\phi_1\rangle\equiv\phi^{i_1\ldots i_s}
\zeta^{[i_1}\alpha^{i_2}\ldots\alpha^{i_s]}|0\rangle\,,
\\
\label{phi0as}
&&
{}~\hspace{-1.5cm}
|\phi_0\rangle\equiv\phi^{i_1\ldots i_{s-1}}
\Bigl((1-\zeta^2)\alpha^{i_1}\ldots\alpha^{i_{s-1}}
+2\sum_{k=1}^{s-1}\alpha^{i_1}\ldots\alpha^{i_{k-1}}
\zeta\alpha\zeta^{i_k}\alpha^{i_{k+1}}\ldots\alpha^{i_{s-1}}
\Bigr)|0\rangle\,.
\end{eqnarray}
As before the $|\phi_1\rangle$ and $|\phi_0\rangle$  transform into one
another under the action of spin operator $M^{IJ}$. The vector
(\ref{pp1p0}) has spin degrees of freedom appropriate to describe totally
antisymmetric massless field. Decomposition (\ref{pp1p0}) reflects the
fact the antisymmetric rank $s$ tensor of $so(d-2)$ algebra can be
decomposed into rank $s$ antisymmetric tensor of $so(d-3)$ algebra and the
ones of rank $s-1$. Let us inspect of action of AdS mass operator on
$|\phi\rangle$. Straightforward calculation gives

$$
A|\phi\rangle=\kappa_1|\phi_1\rangle
+\kappa_2|\phi_0\rangle
+2(2s-1-d-c^\prime)|V\rangle\,,
\qquad
|V\rangle\equiv\phi^{i_1\ldots i_{s-1}}
\alpha^{i_1}\ldots\alpha^{i_{s-1}}|0\rangle\,.
$$
Again the vector $|V\rangle$ does not belong to invariant subspace given in
(\ref{pp1p0}) and therefore we should cancel its contribution. To cancel
the contribution of $|V\rangle$ we set the coefficient in front of this
vector equal to zero and this leads to the solution
$c^\prime=2s-d-1$. Taking into account (\ref{cpr}),
(\ref{ez}) we get then $E_0=d-s-1$. This is lowest energy value for
massless totally antisymmetric field in AdS space-time
(\cite{metsit1},\cite{metsit2}).

{\it Totally symmetric spin $s$ field}. Above analysis can be immediately
extended to arbitrary spin $s$ totally symmetric field. In this case we
have $\lambda=s$ and ${\bf h}=0$ and we start with wavefunction\footnote{
In this case the space $\Lambda$ is given by (\ref{massta}) where we
set $\lambda=s$, $|{\bf h}\rangle=1$ and $m^{ij}=0$.}

\begin{equation}\label{symsphi}
|\phi\rangle=|\phi_s\rangle+(1-\zeta^2)|\phi_{s-1}\rangle+\ldots
\end{equation}
where
\begin{equation}\label{zetdep}
|\phi_s\rangle\equiv
\zeta^{i_1}\ldots\zeta^{i_s}\phi^{i_1\ldots i_s}\,,
\qquad
|\phi_{s-1}\rangle
\equiv
\zeta^{i_1}\ldots\zeta^{i_{s-1}}\phi^{i_1\ldots i_{s-1}}\,.
\end{equation}
Complete expression of $|\phi\rangle$ is given in Appendix B
(see (\ref{sodec})).  Here we exploit first two terms in expansion of
(\ref{sodec}), put there $\rho=1$ and use simplified normalization given
in (\ref{symsphi}).  The dots in (\ref{symsphi}) indicate the subleading
terms which can be read from (\ref{sodec}).  In this case one has

$$
A|\phi\rangle=\kappa_1|\phi_s\rangle
+\kappa_2(1-\zeta^2)|\phi_{s-1}\rangle
+2(5-4s-d-c^\prime)|V\rangle+\ldots\,,
$$
where vector $|V\rangle$ is given by
$$
|V\rangle=\zeta^{i_1}\ldots\zeta^{i_{s-1}}\phi^{i_1\ldots i_{s-1}}\,.
$$
Because the vector $|V\rangle$ again does not belong to the space of
$|\phi\rangle$ given in (\ref{symsphi}) we set factor in front of
$|V\rangle$ equal to zero.  This can be achieved by choosing
$c^\prime=5-d-4s$. Then from (\ref{cpr}), (\ref{ez}) we get value
of $E_0$

$$
E_0=s+d-3\,,
$$
which is nothing but lowest energy value for totally symmetric massless
fields in AdS space-time \cite{metsit1}.

The lowest energy value for massless arbitrary type symmetry representation
has been found in \cite{metsit2} and is given by

$$
E_0=\lambda-k-2+d\,,
$$
where $k$ is determined by the relation
$\lambda=h_1=\ldots h_{k-1}>h_k$. It is for this value $E_0$ that
there is invariant subspace in $\Lambda$ appropriate to describe
massless fields. Because detailed description of that invariant subspace
is too involved, we hope to study it in future publications.

\newsection{Light-cone form of conformal field theory}

In this section we present light-cone reformulation of (free)
 conformal field
theory. The reason for doing  this is that we are going to establish
AdS/CFT correspondence between bulk massless fields and conformal
field theory operators. In the previous sections the bulk massless fields
have been studied within the framework of the light-cone formalism.
Therefore most adequate form for comparison is the light-cone form
of conformal field theory.

\subsection{Lorentz covariant form of conformal field theory}

To keep our presentation as simple as possible we restrict our attention
to the case of arbitrary spin totally symmetric operators that have
canonical conformal dimension given below in (\ref{candim})\footnote{
We do not discuss Lorentz covariant formulation for shadow operators
which have conformal dimension  $\tilde{\Delta}=2-s$. In Appendix C we give
directly light-cone formulation of such operators.}. In this section we
recall main facts of conformal field theory about these operators.  The
reason for this is that we use formulation in terms of generating
functions and the resulting formulas look different from the ones that can
be found in the standard literature on this subject (for instance see
\cite{frpal} and reference therein).

In this section the $so(d-1,2)$ algebra is considered as
algebra of conformal transformations of $(d-1)$-dimensional Minkowski
space-time.  We are interested in spin $s$ totally symmetric
conformal operators

$$
{\cal O}^{a_1\ldots a_s}(x)
$$
that have canonical conformal dimension
\footnote{The fact that expression in r.h.s. of (\ref{candim}) is nothing
but the lowest energy value of spin $s$ massless fields propagating in
$d$ dimensional AdS
space-time has been demonstrated in \cite{metsit1}.}

\begin{equation}\label{candim}
\Delta=s+d-3\,.
\end{equation}
These operators, by definition, are traceless and divergence free

$$
{\cal O}^{aaa_3\ldots a_s}=0\,,
\qquad
\partial_{x^a}{\cal O}^{aa_2\ldots a_s}=0\,.
$$
As above to simplify our presentation we consider Fock space vector
(generating function)

\begin{equation}\label{oriope}
|{\cal O}_{cov}\rangle
\equiv {\cal O}^{a_1\ldots a_s}\alpha^{a_1}\ldots\alpha^{a_s}|0\rangle\,.
\end{equation}
In terms of generating function the traceless and divergence free
conditions take the following form

\begin{eqnarray}
\label{cfttra1}
&&
\bar{\alpha}^a\bar{\alpha}^a|{\cal O}_{cov}\rangle=0\,,
\\
\label{cftdiv1}
&&
\bar{\alpha}^a\partial_{x^a}|{\cal O}_{cov}\rangle=0\,.
\end{eqnarray}
Realization of conformal algebra generators on the space of
operators $|{\cal O}_{cov}\rangle$ is given by

\begin{eqnarray}
\label{covpa}
&&
P^a= \partial^a\,,
\\
\label{covjab}
&&
J^{ab}=x^a \partial^b-x^b\partial^a+M^{ab}\,,
\\
\label{covjd}
&&
D=x^a\partial_{x^a}+\Delta\,,
\\
\label{covka}
&&
K^a=-\frac{1}{2}x_b^2\partial^a+x^a(x^b\partial_{x^b}+\Delta)
+M^{ab}x^b\,,
\end{eqnarray}
where $\partial^a\equiv\eta^{ab}\partial_{x^b}$ and the $so(d-2,1)$ algebra
spin operator $M^{ab}$ is given by

\begin{equation}
\label{lorspiopecft}
M^{ab}=\alpha^a\bar{\alpha}^b-\alpha^b\bar{\alpha}^a\,.
\end{equation}
The condition that the conformal operator $|{\cal O}_{cov}\rangle$ subject
to the constraints (\ref{cfttra1}),(\ref{cftdiv1}) should have the canonical
conformal dimension (\ref{candim}) amounts to the requirement that this
operator constitutes a invariant subspace in representation of conformal
algebra. Indeed the operator $\bar{\alpha}_a^2$ obviously commutes with all
generators of conformal algebra. Operator
$\bar{\alpha}^a\partial_{x^a}$
commutes with the generators $P^a$, $D$ and $J^{ab}$ on the space of
$|{\cal O}_{cov}\rangle$.  As to commutation relation of
$\bar{\alpha}^a\partial_{x^a}$ with $K^a$ we get

$$
[\bar{\alpha}^b\partial_{x^b},K^a]=\bar{\alpha}^a
(\Delta-\alpha^b\bar{\alpha}^b-d+3)
+x^a\bar{\alpha}^b\partial_{x^b}+\alpha^a\bar{\alpha}_b^2\,.
$$
Making use of this and the constraints (\ref{cfttra1}), (\ref{cftdiv1}) we
find

$$
[\bar{\alpha}\partial,K^a]|{\cal O}_{cov}\rangle
=\bar{\alpha}^a
(\Delta-s-d+3)|{\cal O}_{cov}\rangle\,.
$$
From this it clear that only for canonical conformal dimension
(\ref{candim}) the operator $|{\cal O}_{cov}\rangle$
subject to the constraints (\ref{cfttra1}),(\ref{cftdiv1})
constitutes the invariant subspace in representation of conformal algebra.

\subsection{Light-cone form of conformal field theory}

Our derivation of light-cone form of generators of conformal algebra
proceeds as follows. Recall that in the bulk the $so(d-1,2)$
algebra was realized on the space of unconstrained physical fields. On
the CFT side our operators are subject to the divergence constraint
(\ref{cftdiv1}).  It is reasonable to solve this constraint and formulate
boundary conformal theory also in terms of unconstrained operators.
Solution to the constraint (\ref{cftdiv1}) is easily found to be

\begin{equation}\label{oo1}
|{\cal O}_{cov}(\alpha^+,\alpha^-,\alpha^i)\rangle
=\exp\Bigl(-\frac{\alpha^+}{\partial^+}(\bar{\alpha}^+\partial^-
+\bar{\alpha}^i\partial^i)\Bigr)
|{\cal O}(\alpha^-,\alpha^i)\rangle^\prime
\end{equation}
A traceless operator $|{\cal O}\rangle^\prime$ is an unconstrained
operator. As compared to the original operator $|{\cal O}_{cov}\rangle$ this
$|{\cal O}\rangle^\prime$ does not depend on oscillator $\alpha^+$.  The
second step in our derivation is to choose a new basis in which the
generators

\begin{equation}\label{simgen}
P^a,\,
\qquad
J^{+i}\,,
\qquad
J^{+-}\,,
\qquad
K^+\,,
\end{equation}
take form as simple as possible. Namely we choose a basis in which
the generators given in (\ref{simgen}) do not depends on matrix $M^{ab}$.
This can be done step by step. First we choose a basis
which makes $J^{+i}$ independent of $M^{+i}$. Next we choose a basis which
makes $J^{+-}$ independent of $M^{+-}$ and finally we choose a basis
which makes $K^+$ independent of $M^{ab}$.  Details can be found in
Appendix C. Let us present the  final light-cone form of
the generators realized on
conformal theory operators

\begin{eqnarray}
\label{cftpa}
&&
P^a=\partial^a\,,
\\
\label{cftjpi}
&&
J^{+i}=x^+\partial^i-x^i\partial^+\,,
\\
\label{cftjpm}
&&
J^{+-}=x^+\partial^--x^-\partial^+\,,
\\
\label{cftjij}
&&
J^{ij}=x^i\partial_{x^j}-x^j\partial_{x^i}+M^{ij}\,,
\\
\label{cftkp}
&&
K^+=-\frac{1}{2}(2x^+x^- + x_i^2)\partial^+
+x^+D\,,
\\
\label{cftd}
&&
D=x^+\partial^-+x^-\partial^++x^i\partial_{x^i}
+\hat{\Delta}\,,
\\
\label{cftjmi}
&&
J^{-i}=x^-\partial^i-x^i\partial^-
+M^{ij}\frac{\partial^j}{\partial^+}
-\frac{1}{\partial^+}M^i\,,
\end{eqnarray}
where the spin operator $M^{ij}$ is the same as in (\ref{lorspiopecft}).
The generator $M^i$ transforms in vector representation of the spin
operator $M^{ij}$ and satisfies the commutation relations

$$
[M^i,M^{jk}]=\delta^{ij}M^k-\delta^{ik}M^j\,,
\qquad
[M^i,M^i]=\Box M^{ij}\,,
$$
where $\Box$ is the Dalamber operator in $(d-1)$ dimensional
Minkowski space-time

$$
\Box\equiv \partial_{x^a}^2\,.
$$
We use the following realization of unconstrained generating function
$|{\cal O}\rangle$. We decompose $|{\cal O}\rangle$ into irreducible
representations of $so(d-3)$ algebra

\begin{equation}\label{sodec1}
|{\cal O}\rangle
=\left\{
\begin{array}{l}
\sum_{s^\prime}\oplus |{\cal O}^{(1)}_{s^\prime}\rangle
\qquad s^\prime=s,s-2,s-4,\ldots, s-2[\frac{s}{2}]\,,
\\ [7pt]
\sum_{s^\prime}\oplus |{\cal O}^{(2)}_{s^\prime}\rangle
\qquad s^\prime=s-1,s-3,s-5,\ldots, s-2[\frac{s-1}{2}]\,.
\end{array}\right.
\end{equation}
Now a representation of spin part $\hat{\Delta}$ of the
dilatation operator  $D$ (\ref{cftd}) and the operator $M^i$ on conformal
operators $|{\cal O}_{s^\prime}^{(1,2)}\rangle$ is determined to be

\begin{equation}\label{cftdsp}
\hat{\Delta}\equiv\alpha^i\bar{\alpha}^i+d-3\,,
\end{equation}

\begin{eqnarray}
\label{mirep1}
&&
M^i|{\cal O}_{s^\prime}^{(1)}\rangle
=\Box(\alpha^i-\frac{\alpha_j^2\bar{\alpha}^i}{2s^\prime+d-7})
|{\cal O}_{s^\prime-1}^{(2)}\rangle
+\frac{\bar{\alpha}^i}{2s^\prime+d-3}|{\cal O}_{s^\prime+1}^{(2)}\rangle
\\
\label{mirep2}
&&
M^i|{\cal O}_{s^\prime}^{(2)}\rangle
=\Box a(s,s^\prime)
(\alpha^i-\frac{\alpha_j^2\bar{\alpha}^i}{2s^\prime+d-7})
|{\cal O}_{s^\prime-1}^{(1)}\rangle
+\frac{a(s,s^\prime+1)}{2s^\prime+d-3}
\bar{\alpha}^i|{\cal O}_{s^\prime+1}^{(1)}\rangle
\end{eqnarray}
where

$$
a(s,s^\prime)\equiv (s-s^\prime+1)(s+s^\prime+d-5)\,.
$$
Light-cone form of generators for shadow operator can be obtained
from (\ref{cftpa})-(\ref{mirep2}) by making there the following
substitutions: i) the operator $\hat{\Delta}$ in
(\ref{cftkp}), (\ref{cftd}) should be replaced by
$\tilde{\hat{\Delta}}=2-\alpha^i\bar{\alpha}^i$;
ii) in equations (\ref{mirep1}), (\ref{mirep2}) the operator
$\Box$ should be deleted in front of
$|{\cal O}_{s^\prime-1}^{(1,2)}\rangle$
and should be placed in front of $|{\cal O}_{s^\prime+1}^{(1,2)}\rangle$.

\newsection{Light-cone form of AdS/CFT correspondence}

After we have derived the light-cone formulation for  both the bulk
fields and the boundary conformal theory operators we are ready to
demonstrate explicitly
AdS/CFT correspondence. Euclidean version of this correspondence for
various particular cases has been studied in
\cite{Aref'eva:1998nn}--\cite{Polishchuk:1999nh}.
Intertwining operator realization of AdS/CFT correspondence was
investigated in \cite{dobrev}. For review and complete list of references
see \cite{review}.  Here we study correspondence for {\it Lorentzian}
signature of AdS space-time.  As far as we know,
in   this case the correspondence
was discussed only for the case of the scalar field
\cite{Balasubramanian:1999sn}.
We establish correspondence for totally symmetric arbitrary spin massless
fields.  Namely, we match normalizable modes of solution to bulk equations
of motion for massless fields and conformal operators with canonical
conformal dimension. Also we make explicit map between non normalizable
modes and shadow operators which are conformal partners of operators with
canonical conformal dimensions.

What is required is to demonstrate that representation of the $so(d-1,2)$
algebra on bulk physical fields coincides exactly with one for boundary
unconstrained operators.  Also,  it is necessary to make sure that on
both sides one has the same number
of  independent  spin degrees of freedom.

As to spin degrees of freedom,  the matching is straightforward.
Indeed in $d$ - dimensional AdS space-time the massless
totally symmetric field has the following
number of physical spin degrees of freedom

$$
N_{d.o.f}=(2s+d-4)\frac{(s+d-5)!}{(d-4)!s!}\,.
$$
At the same time it is well known  that this $N_{d.o.f}$ is nothing but the
number of independent components of traceless and divergence free operator
in $(d-1)$ dimensional space-time. Thus there is  the same number of spin
degrees of freedom on the both  sides.

Now let us make a comparison of generators for bulk fields and boundary
operators. Important technical simplification is that it is sufficient to
make comparison only for part of algebra spanned by generators

$$
P^a\,,
\qquad
J^{+-}\,,
\qquad
J^{+i}\,,
\qquad
J^{ij}\,,
\qquad
J^{-i}\,,
\qquad
D\,,
\qquad
K^+\,.
$$
It is straightforward to see that the remaining generators $K^-$ and $K^i$
are obtainable from commutation relations of the $so(d-1,2)$ algebra. We
start with a comparison of the kinematical generators (\ref{kingen}). As
for  the generators

$$
P^+\,,
\qquad
P^i\,,
\qquad
J^{+i}
$$
they already coincide on both sides (see (\ref{3spi}),(\ref{3spp})
(\ref{3sjpi}) and (\ref{cftpa}), (\ref{cftjpi})).  Before matching
generators $K^+$ let us consider the dilatation generators $D$. Here we
need explicit form of solution to bulk theory equations of motion. To this
end we cast the equations of motion for totally symmetric field
(\ref{symeqmot5}) into the following form

$$
(-\partial_z^2+\frac{1}{z^2}(-\frac{1}{2}M_{ij}^2
+\frac{(d-4)(d-6)}{4})\Bigr)|\phi\rangle
=\Box |\phi\rangle\,.
$$
Then we decompose  the  field $|\phi\rangle$, which transforms  in
representation of $so(d-2)$ algebra, into irreducible representations of
$so(d-3)$ subalgebra
$|\phi_{s^\prime}\rangle$\footnote{The $|\phi_{s^\prime}\rangle$
for $s^\prime$ given in (\ref{phi1decapp}) and (\ref{phi2decapp})
coincide with $|\phi_{s^\prime}^{(1)}\rangle$ and
$|\phi_{s^\prime}^{(2)}\rangle$ respectively.}

$$
|\phi\rangle=\sum_{s^\prime=0}^s \oplus |\phi_{s^\prime}\rangle\,.
$$
Because of the  relation

$$
\frac{1}{2}M_{ij}^2|\phi_{s^\prime}\rangle
=-s^\prime(s^\prime+d-5)|\phi_{s^\prime}\rangle
$$
the equations of motion for $|\phi_{s^\prime}\rangle$ take the form

$$
(-\partial_z^2+\frac{1}{z^2}(\nu^2-\frac{1}{4}))\phi_{s^\prime}
=\Box \phi_{s^\prime}\,,
\qquad
\nu\equiv s^\prime+\frac{d-5}{2}\,.
$$
Normalizable solutions to these equations are

\begin{equation}\label{norsol}
|\phi_{s^\prime}(x,z)\rangle
=\sqrt{qz}J_{s^\prime+\frac{d-5}{2}}(qz)
q^{-(s^\prime+\frac{d-4}{2})}|{\cal O}_{s^\prime}(x)\rangle\,,
\qquad
q\equiv \sqrt{\Box}\,.
\end{equation}
In the r.h.s. we use the notation $|{\cal O}_{s^\prime}\rangle$ since we
are going to demonstrate that these operators are indeed the conformal
operators. Namely we shall prove that AdS transformations for
$|\phi_{s^\prime}\rangle$ lead to conformal theory transformations for
$|{\cal O}_{s^\prime}\rangle$\footnote{The $|{\cal O}_{s^\prime}\rangle$
for $s^\prime$ given in the first row and the second row of (\ref{sodec1})
coincide with $|{\cal O}_{s^\prime}^{(1)}\rangle$ and
$|{\cal O}_{s^\prime}^{(2)}\rangle$ respectively}.
Asymptotic behavior of the solution above is given by

$$
|\phi_{s^\prime}(x,z)\rangle
\stackrel{z\rightarrow 0}{\longrightarrow}
z^{s^\prime+\frac{d-4}{2}}
|{\cal O}_{s^\prime}(x)\rangle\,.
$$
From this expression and (\ref{padsdm}) it is straightforward to see that

\begin{equation}\label{dadsdcft}
\lim_{z\rightarrow 0}
z^{-s^\prime-\frac{d-4}{2}}D_{ads}|\phi_{s^\prime}\rangle
=D_{cft}|{\cal O}_{s^\prime}\rangle\,.
\end{equation}
Here and below we use the notation $G_{ads}$ and $G_{cft}$ to indicate
realization of $so(d-1,2)$ algebra generators on the  bulk fields
(\ref{3spi})-(\ref{3skm}) and conformal operators
(\ref{cftpa})-(\ref{cftjmi}) respectively.
Thus the operators $D_{ads}$ and $D_{cft}$ also match.
Taking into account the expressions (\ref{3skp}), (\ref{cftkp}) and
(\ref{dadsdcft}) we get immediately

$$
\lim_{z\rightarrow 0}
z^{-s^\prime-\frac{d-4}{2}}K^+_{ads}|\phi_{s^\prime}\rangle
=K^+_{cft}|{\cal O}_{s^\prime}\rangle\,.
$$
The remaining generators of the algebra we need to match
are $P^-$, $J^{+-}$, $J^{-i}$. Let us consider $P^-_{ads}$ and $P_{cft}^-$

$$
P_{ads}^-=-\frac{\partial_J^2}{2\partial^+}
+\frac{1}{2z^2\partial^+}
\Bigl(-\frac{1}{2}M_{ij}^2+\frac{(d-4)(d-6)}{4}\Bigr)\,,
\qquad
P^-_{cft}=\partial^-\,.
$$
In the bulk generator $P_{ads}^-$ given above we cannot send the coordinate
$z$ to zero. The point is that the $P_{ads}^-$ consists of second
derivative with respect to $z$. Therefore the operator $P^-_{ads}$ does
not commute with $z$, $[P^-,z] \sim \partial_z$,
and the r.h.s. is not equal on the representation space of bulk theory.
Note that the bulk generators are defined on space of initial
data, i.e., for $x^+=const$ .
Therefore the fact that we cannot send the coordinate $z$ to
zero in the bulk $P^-$ implies that we cannot directly map
the initial data of the
bulk theory into the boundary conformal theory operators.
The point is that in order to put $z$ equals to zero in bulk generators
we have to replace the representation defined on initial data
by the representation defined on the space of solutions. This implies
simply that we should use the fact that on the space of solutions we
have Schr\"odinger equations of motion $P_{ads}^-\phi=\partial^- \phi$. In
other words,  on the space of solutions we can simply replace $P_{ads}^-$
by $\partial^-$

\begin{equation}\label{padsdm}
P_{ads}^-=\partial^-\,.
\end{equation}
So generators $P_{ads}^-$ and $P_{cft}^-$ also match. Taking this into
account it is straightforward to see that the generators $J_{ads}^{+-}$
(\ref{3sjpm}) and $J^{+-}_{cft}$ (\ref{cftjpm})

$$
J_{ads}^{+-}=x^+P^- -x^-\partial^+\,,
\qquad
J^{+-}_{cft}=x^+\partial^- - x^-\partial^+
$$
also coincide.  The last step is to match the generators
$J_{ads}^{-i}$ and $J_{cft}^{-i}$. To this end we rewrite the
$J_{ads}^{-i}$ given in (\ref{3sjmi}) as follows

\begin{equation}\label{jadsdec}
J_{ads}^{-i}=x^-\partial^i-x^iP_{ads}^-
+M^{ij}\frac{\partial^j}{\partial^+}
-\frac{1}{\partial^+}M_{ads}^i\,,
\qquad
M^i_{ads}=M^{zi}\partial_z+\frac{1}{2z}\{M^{zj},M^{ji}\}\,.
\end{equation}
Using (\ref{padsdm}) and comparing the above expression for
$J_{ads}^{-i}$ with $J_{cft}^{-i}$ given in (\ref{cftjmi}) we
conclude that all that remains to do is to match $M_{ads}^i$ given in
(\ref{jadsdec}) and $M_{cft}^i$ given in  (\ref{mirep1}), (\ref{mirep2}).
Technically, this is the most difficult point of matching. The fact of
coincidence of the operators $M_{ads}^i$ and $M_{cft}^i$ we prove
by direct calculation. The action of $M_{cft}^i$ is given in
(\ref{mirep1}), (\ref{mirep2}). By acting with operator $M^i_{ads}$ on
space of solutions given in (\ref{norsol}) we obtain the representation of
operator  $M_{ads}^i$ in $|{\cal O}_{s^\prime}\rangle$.  We should to
prove that this representation of operator  $M_{ads}^i$ in $|{\cal
O}_{s^\prime}\rangle$ coincides with (\ref{mirep1}) (\ref{mirep2}).  This
fact is proved in Appendix B.

Finally,  let us  write $AdS/CFT$ correspondence for bulk symmetric spin
$s$ massless field and corresponding boundary conformal theory operator.
From (\ref{norsol}) we can read the following relationship

\begin{equation}\label{filope}
\lim_{z\rightarrow 0} z^{-\hat{\Delta}+\Delta_0}
|\phi_{s^\prime}(x,z)\rangle
=|{\cal O}_{s^\prime}(x)\rangle\,,
\qquad
s^\prime =0,1,\ldots, s\,,
\end{equation}
where $\hat{\Delta}$ is a spin part of dilatation generator $D_{cft}$
(\ref{cftdsp}) while $\Delta_0$ is a canonical dimension of bulk massless
field in $d$-dimensional AdS space-time

$$
\Delta_0=\frac{d-2}{2}\,.
$$
Above we matched normalizable solutions of bulk theory equations of motion
and boundary conformal theory operators that have canonical
conformal dimension.

The generalization of our analysis to the case
non-normalizable modes is straightforward. In this case we are going to
demonstrate that {\it non-normalizable  bulk modes correspond to the shadow
operators of boundary conformal field theory}.  To this end let us write
down explicitly the non-normalizable solutions to the bulk equations of
motion\footnote{To keep discussion from becoming unwieldy
here we restrict our attention to even $d$. In this case the solutions
given in (\ref{norsol}) and (\ref{nonnorsol}) are independent.}

\begin{equation}\label{nonnorsol}
|\phi_{s^\prime}(x,z)\rangle_{non-norm}
=\sqrt{qz}J_{-s^\prime-\frac{d-5}{2}}(qz)
q^{s^\prime+\frac{d-6}{2}}|\tilde{{\cal O}}_{s^\prime}(x)\rangle\,.
\end{equation}
They have the following asymptotic behavior

$$
|\phi_{s^\prime}(x,z)\rangle_{non-norm}
\stackrel{z\rightarrow 0}{\longrightarrow}
z^{-s^\prime-\frac{d-6}{2}}
|\tilde{{\cal O}}_{s^\prime}(x)\rangle\,.
$$
Now the fact that non-normalizable solution indeed corresponds to shadow
operator follows from the relation

\begin{equation}\label{}
\lim_{z\rightarrow 0}z^{s^\prime+\frac{d-6}{2}}D_{ads}
|\phi_{s^\prime}\rangle_{non-norm}
=\tilde{D}_{cft}|\tilde{{\cal O}}_{s^\prime}\rangle\,,
\end{equation}
where $\tilde{D}_{cft}$ is the dilatation operator for the
shadow operator

\begin{equation}\label{dshad}
\tilde{D}_{cft}=x^a\partial_{x^a}+\tilde{\hat{\Delta}}\,,
\qquad
\tilde{\hat{\Delta}}=2-\alpha^i\bar{\alpha}^i\,.
\end{equation}
After this the remaining generators can be matched in the  same manner as
it was
done for normalizable modes. Relation between  non normalizable solutions
and shadow operators is given by then by the formulas (\ref{filope}) where
the spin operator $\hat{\Delta}$ should be replaced by the one for shadow
field (\ref{dshad}).

\newsection{Conclusions}

We have developed  the light-cone formalism in AdS space-time.
In this paper we applied this formalism to the  study of
AdS/CFT correspondence.
Because the formalism we presented is algebraic in nature it allows us to
treat  fields with arbitrary spin on equal footing. Comparison of  this
formalism with other approaches   available in the
 literature leads us to  the conclusion
that this is a very efficient formalism.

The results presented here should have a number of interesting applications
and generalizations, some of which are:

(i)
generalization to AdS/CFT correspondence between arbitrary spin massive
fields and related  operators at the boundary;

(ii) generalization to supersymmetry and applications to type IIB
supergravity in $AdS_5\times S^5$ background \cite{metsugra}
and then to strings in this background;

 (iii) extension of light-cone formulation of conformal field
theory to the level of OPE's and study of light-cone form of AdS/CFT
correspondence at the level of correlation functions;

 (iv) application of
light-cone formalism to the study of the S-matrix along the lines of
\cite{polch}--\cite{gidd};

 (v) applications to interaction vertices for
higher massless spin fields in AdS space-time.

In this paper we have discussed AdS/CFT correspondence between
massless arbitrary spin fields in AdS space-time and conformal arbitrary
spin operators at boundary at the level of free equations of motion. By
now it is known that to construct self-consistent interaction of massless
higher spin fields it is necessary to introduce, among other things,  a
infinite chain of anti-de Sitter massless fields which consists of every spin
just once \cite{vas1}. This implies that to maintain AdS/CFT correspondence
for such interaction equations of motion we should also introduce an
infinite chain of conformal operators at the boundary. In this respect it
would be interesting to extend the analysis of this paper to the case of
that infinite chain of interacting massless fields and corresponding conformal
operators.

We strongly believe that the
light-cone formalism developed in this paper will
be useful for better  understanding  of strings in AdS/RR-charge
 backgrounds.

\begin{center}
{\bf Acknowledgments}
\end{center}

I would like to thank A. Tseytlin for reading the manuscript and
comments.
This  work was supported in part
by INTAS grant No.96-538 and the Russian Foundation for Basic Research
Grant No.99-02-17916.

\setcounter{section}{0}
\setcounter{subsection}{0}

\appendix{Gauge invariant equations of motion for totally
symmetric fields}

In order to find gauge invariant equations of motion we use the
algebra of commutation relations for operators that can be constructed by
using the commuting oscillators $\alpha^A$ (\ref{comosc}) and Lorentz
covariant derivative $D^A$.  Starting with the commutator

$$
[\hat{\partial}_A,\hat{\partial}_B]=\Omega_{AB}{}^C\hat{\partial}_C\,,
\qquad
\Omega^{ABC}\equiv -\omega^{ABC}+\omega^{BAC}\,,
\qquad
\omega_A{}^{BC}\equiv e^\mu_A\omega_\mu^{BC} \,,
$$
where $\Omega^{ABC}$ is a contorsion tensor we get immediately the
following basic commutation relation

\begin{equation}\label{dadbap2}
[D_A,D_B]=\Omega_{AB}{}^CD_C+\frac{1}{2}R_{AB}^{CE}M^{CE}\,.
\end{equation}
Multiplying by oscillators both sides of (\ref{dadbap2})
and taking into account the commutators

$$
[\alpha^A,D_B]=\omega_B{}^{AC}\alpha^C\,,
\qquad
[\bar{\alpha}^A,D_B]=\omega_B{}^{AC}\bar{\alpha}^C\,,
$$
we get the following useful commutators

$$
[\bar{\alpha} D,\alpha D]=D_A^2+\omega^{AAB}D_B
-\frac{1}{4}R_{AB}^{CD}M^{CD}M^{AB}\,,
$$
where $R^{AB}_{CD}$ is a Rieman tensor in tangent space

$$
R^{AB}_{CD}
=\hat{\partial}_C\omega_D{}^{AB}
+\omega_C{}^{AE}\omega_D{}^{EB}
+\omega_{CD}{}^E\omega^{EAB}-(C\leftrightarrow D)\,.
$$
Next important commutator is given by

\begin{eqnarray}
[[\bar{\alpha}D,\alpha D],\alpha D]
&=&2
R_{AB}\alpha^A D^B-2R_{AB}^{DE}\alpha^AM^{DE}D_B
\nonumber\\
\label{badadad}
&+&
\frac{1}{4}D_CR_{AB}^{DE}\alpha^CM_{AB}M^{DE}
+D_BR_{CA}\alpha^AM_{BC}\,,
\end{eqnarray}
where $R^A_B=R^{CA}_{CB}$ is a Richi tensor in tangent space. Note that in
this appendix and only in this appendix the $R^{AB}$ indicates the Richi
tensor. In the remainder of this paper the $R^{AB}$ is used as it is defined
in (\ref{rab}). To derive (\ref{badadad}) one can use the following
useful commutation relations

$$
[D_A^2+\omega^{AAB}D_B,\alpha D]=R_{AB}\alpha^A D^B
+R_{AB}^{CD}\alpha^BM^{CD}D^A+D_CR_{DB}\alpha^BM^{CD}\,,
$$
$$
[\omega^{AAB}D_B,\alpha D]
=(-\hat{\partial}_B\omega^{AAC}+\omega^{AAD}\omega^{BDC})\alpha^BD_C
+\frac{1}{2}\omega^{AAB}R_{BC}^{DE}\alpha^CM^{DE}\,.
$$
Note that all abovemenioned commutation relations are valid for
space-time of arbitrary geometry. Due to that these relations might be
useful for a number of interesting applications to gauge
invariant equations of motion for totally symmetric fields propagating in
space-times of arbitrary geometry. For the case of AdS geometry the Rieman
tensor satisfies the relationships

$$
R^{ABCD}=-\eta^{AC}\eta^{BD}+\eta^{AD}\eta^{BC}\,,
\qquad
D_ER^{AB}_{CD}=0\,.
$$
With the above algebra of commutators at hand the derivation of
the gauge invariant equations of motion is relatively straightforward.
We look for equations of motion of the form $L|\Phi\rangle=0$ where $L$
is a polynomial  of the second order in $\alpha D$ and $\bar{\alpha}D$ and
impose the following conditions. i) Operator $L$ commutes with
oscillator number operator $\alpha\bar{\alpha}$. ii) On the space of
double traceless tensor the operator $L$ commutes with operator
$(\bar{\alpha}^2)^2$. This condition amount to the requirement that the
equations of motion and double traceless condition respect each other;
From this requirement we learn that the operator $L$ does not consist of
terms $(\bar{\alpha}^2)^n$, $n>1$. iii) Operator $L$ commutes with gauge
transformation. This implies that equations of motion should be
gauge invariant. Since we use gauge transformation of the form
(\ref{dfa1}) we should impose the condition $[L,\alpha
D]|\Lambda\rangle=0$, where the gauge parameter transformation
$|\Lambda\rangle$ satisfies the traceless constraint (\ref{lamtracon1}).
Making use of commutation relation above one can make sure that the
unique operator $L$ that satisfies these requirement is that given in
(\ref{symeqmot1}).

In the rest of this appendix let us write down some important formulas
we use in this paper. In Poincar\'e coordinates the Lorentz covariant
derivative takes the form

$$
D^A=\hat{\partial}^A+\alpha^z\bar{\alpha}^A
-\alpha^A\bar{\alpha}^z\,.
$$
From this we get the following useful representations

$$
\alpha D=\alpha\hat{\partial}+\alpha^z\alpha\bar{\alpha}
-\alpha^2\bar{\alpha}^z\,,
\qquad
\bar{\alpha}D=\bar{\alpha}\hat{\partial}
+(2-d)\bar{\alpha}^z-\bar{\alpha}^z\alpha\bar{\alpha}
+\alpha^z\bar{\alpha}^2\,,
$$
$$
D_A^2=(\hat{\partial}^A-\alpha^A\bar{\alpha}^z)^2
+2\alpha^z\bar{\alpha}D-\alpha^{z2}\bar{\alpha}^2
+(d-1)\alpha^z\bar{\alpha}^z-\alpha\bar{\alpha}\,.
$$
Some other commutation relations in Poincar\'e coordinates are

$$
[\bar{\alpha}^A,\alpha D]
=\hat{\partial}^A+\alpha^z\bar{\alpha}^A
+\delta_z^A\alpha\bar{\alpha}-2\alpha^A\bar{\alpha}^z\,,
\qquad
[\bar{\alpha}^A, \bar{\alpha} D]=\delta_z^A\bar{\alpha}^2
-\bar{\alpha}^z\bar{\alpha}^A\,,
$$
$$
[\bar{\alpha}^A, D^2]
=2\delta_z^A\bar{\alpha}D-2\bar{\alpha}^z D^A
+(d-2)\delta_z^A \bar{\alpha}^z+\bar{\alpha}^A\,.
$$
These commutation relations are useful while studying the  constraints
that follow from equations of motion in the light-cone gauge.

\appendix{Various forms of realization of $so(d-2)$ algebra
representations}

{\it Stereographic form}. Here we wish to demonstrate that spin operator
given in (\ref{mijgt}) comes from description of $so(d-2)$ algebra
representation in stereographic coordinates. To this end we consider
the simplest case of totally symmetric representations.
Let $\phi^{I_1\ldots I_s}$ be a totally symmetric traceless tensor
field which realizes irreducible representation of $so(d-2)$ algebra.
Introduce vector $a^I$ and consider a generating function

\begin{equation}\label{appphi}
|\phi\rangle =\phi^{I_1\ldots I_s}a^{I_1}\ldots a^{I_s}\,.
\end{equation}
The spin operator in this representation takes the form

$$
M^{IJ}=a^I\partial_{a^J}-a^J\partial_{a^J}\,.
$$
By definition the $|\phi\rangle$ satisfies the constraints

\begin{equation}\label{con12}
a^I\partial_{a^I}|\phi\rangle=s|\phi\rangle\,,
\qquad
\partial_{a^I}^2|\phi\rangle=0\,.
\end{equation}
The first constraint tells us that $|\phi\rangle$ is a monomial degree $s$
in $a^I$ while the second one is a traceless condition.
Stereographic coordinates $\zeta^i$ are defined by the relations

\begin{equation}\label{stecoo}
a^i=2\zeta^i\rho\,,
\qquad
a^z=(1-\zeta^2)\rho\,,
\end{equation}
where $\rho$ is a scale parameter. Making use of chain rules
$$
\partial_{a^i}=\frac{1}{2\rho}\partial_{\zeta^i}
+\frac{\zeta^i}{1+\zeta^2}(\partial_\rho
-\frac{1}{\rho}\zeta\partial_\zeta)\,,
\quad
\partial_{a^z}=\frac{1}{1+\zeta^2}(\partial_\rho
-\frac{1}{\rho}\zeta\partial_\zeta)\,,
\quad
\zeta^2\equiv \zeta^i\zeta^i\,,
\quad
\zeta\partial_\zeta\equiv \zeta^i\partial_{\zeta^i}
$$
one can verify that the constraints (\ref{con12}) and spin operator
$M^{IJ}$ take the following form

\begin{equation}\label{apcon1}
\rho\partial_\rho|\phi\rangle=s|\phi\rangle\,,
\qquad
\Bigl((1+\zeta^2)\partial_{\zeta^i}^2-4(s+\frac{d-5}{2})(\zeta\partial_\zeta
-s)\Bigr)|\phi\rangle=0\,.
\end{equation}
\begin{equation}
\label{mijap}
M^{ij}=\zeta^i\partial_{\zeta^j}-\zeta^j\partial_{\zeta^i}\,,
\qquad
M^{zi}=\frac{1}{2}(1-\zeta^2)\partial_{\zeta^i}
+\zeta^i(\zeta\partial_\zeta-\rho\partial_\rho)\,.
\end{equation}
The operator $\rho\partial_\rho$ commutes with above spin operator
$M^{IJ}$. Due to that in expression for $M^{zi}$ (\ref{mijap}) we can
replace the operator $\rho\partial_\rho$ by its eigenvalue $s$.
After this
comparing the spin operators (\ref{mijap})  with the ones
of group theoretic approach (\ref{mijgt}) we conclude that the
group theoretic approach  indeed leads to the stereographic form of
realization of spin degrees of freedom. Note that by inserting
(\ref{stecoo}) into (\ref{appphi}) the expression for $|\phi\rangle$ can
be cast into the well known textbook form

\begin{equation}\label{sodec}
|\phi(a^I)\rangle
=\rho^s\sum_{s^\prime=0}^s
(1+\zeta^2)^{s-s^\prime}C_{s-s^\prime}^{\frac{d-4}{2}+s^\prime}(t)
|\phi_{s^\prime}(\zeta^i)\rangle\,,
\quad
\zeta\partial_\zeta|\phi_{s^\prime}\rangle=
s^\prime|\phi_{s^\prime}\rangle\,,
\quad
\partial_{\zeta^i}^2|\phi_{s^\prime}\rangle=0\,,
\end{equation}
where $t\equiv(1-\zeta^2)/(1+\zeta^2)$ and $C_k^\alpha$ is a
Gegenbauer polynom. The $|\phi_{s^\prime}\rangle$ is totally symmetric
traceless rank $s^\prime$ tensor of $so(d-3)$ algebra.  The formula
(\ref{sodec}) is a decomposition of irreducible representation of
$so(d-2)$ algebra into ones of $so(d-3)$ algebra.

{\it Simple form}. Here we wish to describe an form of
realization of totally symmetric $so(d-2)$ algebra representation in
terms of $so(d-3)$ which as compared to stereographic form does not
involve special functions like Gegenbauer polynoms and which we did not
find in standard literature on this subject.  This representation turns
out to be more convenient for establishing AdS/CFT correspondence.  As
before we focus on symmetric field. Let $\phi^{I_1\ldots I_s}$ be a
totally symmetric traceless tensor field.  Consider a generating function

$$
|\phi\rangle\equiv \phi^{I_1\ldots I_s}\alpha^{I_1}\ldots
\alpha^{I_s}|0\rangle\,,
$$
which satisfies the constraint
\begin{equation}
\label{trco}
\alpha^I\bar{\alpha}^I|\phi\rangle=s|\phi\rangle\,,
\qquad
\bar{\alpha}^I\bar{\alpha}^I|\phi\rangle=0\,.
\end{equation}
One can consider the second constraint in (\ref{trco}) as the second
order differential equation with respect to oscillator variable $\alpha^z$

$$
(\bar{\alpha}^{z2}+\omega^2)|\phi(\alpha^z,\alpha^i)\rangle=0\,,
\qquad
\omega^2=\bar{\alpha}^i\bar{\alpha}^i\,.
$$
Obvious solution to this equation is found to be
\begin{equation}\label{pp1p2}
|\phi(\alpha^z,\alpha^i)\rangle
=\cos (\omega \alpha^z)|\phi_s(\alpha^i)\rangle
+\frac{\sin (\omega \alpha^z)}{\omega}|\phi_{s-1}(\alpha^i)\rangle\,,
\end{equation}
where $|\phi_s\rangle$ and $|\phi_{s-1}\rangle$
rank $s$ and $s-1$ traceful tensors, i.e. they are reducible
representations of the $so(d-3)$ algebra. The solution (\ref{pp1p2})
reflects well know fact that symmetric traceless rank $s$  tensor of
$so(d-2)$ algebra can be decomposed into symmetric rank $s$ and $s-1$
traceful tensors of $so(d-3)$ algebra. These tensors satisfy the
constraints

$$
\alpha^i\bar{\alpha}^i|\phi_s\rangle=s|\phi_s\rangle\,,
\qquad
\alpha^i\bar{\alpha}^i|\phi_{s-1}\rangle=(s-1)|\phi_{s-1}\rangle\,,
\qquad
\bar{\alpha}^z |\phi_{s,s-1}\rangle=0\,.
$$
Straightforward calculations gives
\begin{equation}\label{mzinew}
M^{zi}|\phi\rangle
=\cos(\omega \alpha^z)(-\alpha^i)|\phi_{s-1}\rangle
+\frac{\sin (\omega \alpha^z)}{\omega}
(\bar{\alpha}^i+\alpha^i\omega^2)|\phi_s\rangle\,.
\end{equation}
Instead of the representation (\ref{pp1p2}) we express
$|\phi\rangle$ in terms $|\phi_{s,s-1}\rangle$ as follows

\begin{equation}\label{phph1ph2app}
|\phi\rangle
=\left(
\begin{array}{l}
|\phi_s\rangle
\\[5pt]
|\phi_{s-1}\rangle
\end{array}
\right)\,.
\end{equation}
Then from the (\ref{mzinew}) we get immediately the
following representation for spin operator $M^{zi}$ on the generating
function $|\phi\rangle$

\begin{equation}\label{mziapp}
M^{zi}=\sigma_-(\bar{\alpha}^i
+\alpha^i\bar{\alpha}_j^2)-\sigma_+ \alpha^i\,,
\qquad
\sigma_-\equiv\left(
\begin{array}{ll}
0 & 0\\
1 &0
\end{array}\right)\,,
\qquad
\sigma_+\equiv\left(
\begin{array}{ll}
0 & 1\\
0 &0
\end{array}\right)\,.
\end{equation}
The representation of $M^{ij}$ on $|\phi\rangle$ has the usual form
\begin{equation}\label{mijapp}
M^{ij}=\alpha^i\bar{\alpha}^j-\alpha^j\bar{\alpha}^i\,.
\end{equation}
For comparison with boundary theory it is convenient to exploit the
following realization of $|\phi_{s,s-1}\rangle$. Decompose
the reducible generating functions $|\phi_{s,s-1}\rangle$ into irreducible
representations of $so(d-3)$ algebra

\begin{eqnarray}
\label{phi1decapp}
&&\hspace{-1cm}
|\phi_s\rangle=\sum_{s^\prime}
(-\alpha_j^2)^{[\frac{s-s^\prime}{2}]}|\phi_{s^\prime}^{(1)}\rangle\,,
\qquad
s^\prime=s,s-2,s-4,\ldots, s-2[\frac{s}{2}]\,,
\\
\label{phi2decapp}
&&\hspace{-1cm}
|\phi_{s-1}\rangle=\sum_{s^\prime}
(-\alpha_j^2)^{[\frac{s-s^\prime-1}{2}]}|\phi_{s^\prime}^{(2)}\rangle\,,
\quad
s^\prime=s-1,s-3,s-5,\ldots, s-2[\frac{s-1}{2}]\,,
\end{eqnarray}
$(\alpha^i\bar{\alpha}^i-s^\prime)|\phi_{s^\prime}^{(1,2)}\rangle
=\bar{\alpha}_i^2|\phi_{s^\prime}^{(1,2)}\rangle=0$.
In this basis defining the action $M^{zi}$ on
$|\phi_{s^\prime}^{(1,2)}\rangle$ with the relation

$$
M^{zi}|\phi\rangle
=\left(
\begin{array}{l}
\sum_{s^\prime}
(-\alpha_j^2)^{[\frac{s-s^\prime}{2}]}
M^{zi}|\phi_{s^\prime}^{(1)}\rangle
\\[7pt]
\sum_{s^\prime}
(-\alpha_j^2)^{[\frac{s-s^\prime-1}{2}]}
M^{zi}|\phi_{s^\prime}^{(2)}\rangle
\end{array}
\right)
$$
we get
\begin{eqnarray}
\label{appmziphi1}
&&
{}~\hspace{-2cm}
M^{zi}|\phi_{s^\prime}^{(1)}\rangle
=-(\alpha^i-\frac{\alpha_j^2\bar{\alpha}^i}{2s^\prime+d-7})
|\phi_{s^\prime-1}^{(2)}\rangle
+\frac{\bar{\alpha}^i}{2s^\prime+d-3}|\phi_{s^\prime+1}^{(2)}\rangle\,,
\\
\label{appmziphi2}
&&
{}~\hspace{-2cm}
M^{zi}|\phi_{s^\prime}^{(2)}\rangle
=-a(s,s^\prime)(\alpha^i-\frac{\alpha_j^2\bar{\alpha}^i}{2s^\prime+d-7})
|\phi_{s^\prime-1}^{(1)}\rangle
+a(s,s^\prime+1)
\frac{\bar{\alpha}^i}{2s^\prime+d-3}|\phi_{s^\prime+1}^{(1)}\rangle\,.
\end{eqnarray}

{\it Evaluation of action of operator $M^i$}. Here by using above simple
form of realization of $so(d-2)$ algebra representation we
evaluate the action of operator

\begin{equation}\label{miapp}
M^i\equiv M^{zi}\partial_z+\frac{1}{2z}\{M^{zj},M^{ji}\}
\end{equation}
on the space of solution given in (\ref{norsol}). We use representation
for spin operator $M^{IJ}$ given in (\ref{mziapp}), (\ref{mijapp}). For
this representation we derive in a straightforward way

\begin{eqnarray*}
&&
\{M^{zj},M^{ji}\}=\sigma_+\Bigl(\alpha^i(2\alpha_j\bar{\alpha}_j+d-4)
-2\alpha_j^2\bar{\alpha}^i\Bigr)
\\
&&
+\sigma_-\Bigl(
\bar{\alpha}^i(M_{jk}^2+2(\alpha_j\bar{\alpha}_j)^2
+2(d-4)\alpha_j\bar{\alpha}_j+d-6)
-\alpha^i(d+2+2\alpha_j\bar{\alpha}_j)\bar{\alpha}_k^2\Bigr)\,.
\end{eqnarray*}
By acting with this on $|\phi\rangle$ given in (\ref{phph1ph2app})
and using the decompositions (\ref{phi1decapp}), (\ref{phi2decapp})
we get the action of operator $\{M^{zj},M^{ji}\}$ on
$|\phi_{s^\prime}^{(1,2)}\rangle$

\begin{eqnarray*}
&&
\{M^{zj},M^{ji}\}|\phi_{s^\prime}^{(1)}\rangle
=(2s^\prime+d-6)(\alpha^i-\frac{\alpha_j^2\bar{\alpha}^i}{2s^\prime+d-7})
|\phi_{s^\prime-1}^{(2)}\rangle
+\frac{2s^\prime+d-4}{2s^\prime+d-3}\bar{\alpha}^i
|\phi_{s^\prime+1}^{(2)}\rangle\,,
\\
&&
\{M^{zj},M^{ji}\}|\phi_{s^\prime}^{(2)}\rangle
=(2s^\prime+d-6)
a(s,s^\prime)(\alpha^i-\frac{\alpha_j^2\bar{\alpha}^i}{2s^\prime+d-7})
|\phi_{s^\prime-1}^{(1)}\rangle
\\
&&
\hspace{3cm}
+a(s,s^\prime+1)\frac{2s^\prime+d-4}{2s^\prime+d-3}\bar{\alpha}^i
|\phi_{s^\prime+1}^{(1)}\rangle\,,
\end{eqnarray*}
Now we (i) combine these relations with
(\ref{appmziphi1}), (\ref{appmziphi2}) and find action of operator $M^i$
(\ref{miapp}) on $|\phi_{s^\prime}^{(1,2)}\rangle$; (ii) use solutions of
equations of motion for $|\phi_{s^\prime}\rangle$ given in (\ref{norsol});
(iii) exploit the following relationships for Bessel functions

\begin{eqnarray*}
(-\partial_z+\frac{2s^\prime+d-6}{2z})
\sqrt{z}J_{s^\prime+\frac{d-7}{2}}(z)
=
(\partial_z+\frac{2s^\prime+d-4}{2z})
\sqrt{z}J_{s^\prime+\frac{d-3}{2}}(z)
=\sqrt{z}J_{s^\prime+\frac{d-5}{2}}(z)
\end{eqnarray*}
After this we get that the action of $M^i$ on boundary values
$|{\cal O}_{s^\prime}^{(1,2)}\rangle$
coincides with the action of $M^i_{cft}$ given in
(\ref{mirep1}), (\ref{mirep2}).

\appendix{Transformation of conformal theory generators to light-cone form}

In this appendix we describe transformations that take the generators
given in (\ref{covpa})-(\ref{covka}) to the those of light-cone form
given in (\ref{cftpa})-(\ref{cftjmi}).  Again we start our analysis with
kinematical generators (\ref{kingen}).  First we consider the generators
$J^{+i}$ and $J^{+-}$.  The original $J^{+i}$ and $J^{+-}$ acting on
$|{\cal O}_{cov}\rangle$ are defined by (\ref{covjab}). Taking into account
(\ref{oo1}) it is straightforward to see that the generators $J^{+i}$ and
$J^{+-}$ in $|{\cal O}\rangle^\prime$ basis take the form

\begin{equation}\label{o2bas}
J^{+i}=x^+\partial^i-x^i\partial^+ - \alpha^i\bar{\alpha}^+\,,
\qquad
J^{+-}=x^+\partial^--x^-\partial^+
-\alpha^-\bar{\alpha}^+\,.
\end{equation}
Our first step is to find the transformation that cancels out the
oscillator terms in the generator $J^{+i}$. This is achieved by the
following transformation

\begin{equation}\label{o1o2}
|{\cal O}\rangle^\prime
=\exp(\alpha^i\bar{\alpha}^+\frac{\partial^i}{\partial^+})
|{\cal O}\rangle^{\prime\prime}
\end{equation}
In $|{\cal O}\rangle^{\prime\prime}$ basis
the generator $J^{+i}$ takes desired form given in (\ref{cftjpi})
while the generator $J^{+-}$ is not changed (see (\ref{o2bas})).
Because of relations (\ref{oo1}) and (\ref{o1o2}) the original
$|{\cal O}_{cov}\rangle$ and $|{\cal O}\rangle^{\prime\prime}$ are related
as follows

$$
|{\cal O}_{cov}\rangle
=\exp(-\frac{\alpha^+}{\partial^+}(\bar{\alpha}^+\partial^-
+\bar{\alpha}^i\partial^i))
\exp(\alpha^i\bar{\alpha}^+\frac{\partial^i}{\partial^+})
|{\cal O}\rangle^{\prime\prime}\,.
$$
Taking into account the formula
$$
e^{\alpha^iX^i }e^{\bar{\alpha}^iY^i}
=e^{\alpha^i X^i+\bar{\alpha}^iY^i-\frac{1}{2}X^iY^i}\,,
$$
this can be rewritten as
$$
|{\cal O}_{cov}\rangle=e^{-\Gamma}|{\cal O}\rangle^{\prime\prime}\,,
\qquad
\Gamma=\frac{\alpha^+\bar{\alpha}^+}{2\partial^{+2}}\Box
+M^{+i}\frac{\partial^i}{\partial^+}\,.
$$
In order to cancel oscillator term in expression for $J^{+-}$
(\ref{o2bas}) we make the transformation

$$
|{\cal O}\rangle^{\prime\prime}=
(-\partial^+)^{\alpha^-\bar{\alpha}^+}
|{\cal O}\rangle^{\prime\prime\prime}\,.
$$
The original operator $|{\cal O}_{cov}\rangle$ (\ref{oriope}) takes then
the form
$$
|{\cal O}_{cov}\rangle=e^{-\Gamma}(-\partial^+)^{\alpha^-\bar{\alpha}^+}
|{\cal O}\rangle^{\prime\prime\prime}
$$
Thus in $|{\cal O}\rangle^{\prime\prime\prime}$ basis
the generators $J^{+i}$ and $J^{+-}$ take the desired form given
in (\ref{cftjpi}) and (\ref{cftjpm}). Let us now consider the remaining
kinematical generator $K^+$. In $|{\cal O}\rangle^{\prime\prime\prime}$
basis this generator takes the following form

$$
K^+=K_0^++(\Delta-\alpha^-\bar{\alpha}^+)x^+
+\frac{1}{2}\partial^+\alpha_i^2\bar{\alpha}^{+2}\,,
\qquad
K_0^+\equiv -\frac{1}{2}x_i^2\partial^+
+x^+(x^+\partial^-+x^i\partial^i)\,.
$$
Note that in $|{\cal O}\rangle^{\prime\prime\prime}$ basis
the traceless condition (\ref{cfttra1}) takes the
following form

$$
(\bar{\alpha}_i^2-\Box\bar{\alpha}^{+2})
|{\cal O}\rangle^{\prime\prime\prime}=0\,.
$$
Before to proceed we wish to transform this traceless condition to the
form that does not consist of $\Box$.  To this end we introduce $|{\cal
O}\rangle^{iv}$ basis

$$
|{\cal O}\rangle^{\prime\prime\prime}
=\sqrt{\Box}^{\,(-\alpha^-\bar{\alpha}^+)}
|{\cal O}\rangle^{iv}\,.
$$
In the $|{\cal O}\rangle^{iv}$ basis the traceless condition takes desired
form
$$
(\bar{\alpha}_i^2-\bar{\alpha}^{+2})
|{\cal O}\rangle^{iv}=0\,,
$$
while the generators $K^+$, $J^{-i}$ take the form

\begin{eqnarray}
\label{intkp}
&&
K^+=K_0^++\Delta x^+ +\frac{\partial^+}{2\Box}
(s(s+d-5)+\frac{1}{2}M_{ij}^2)\,,
\\
\label{intjmi}
&&
J^{-i}=x^-\partial^i-x^i\partial^-
+M^{ij}\frac{\partial^j}{\partial^+}
-\frac{1}{\partial^+}M^i\,,
\end{eqnarray}
where the operator $M^{ij}$ is given in (\ref{mijapp}), while the
operator $M^i$ is given by
$$
M^i\equiv
\sqrt{\Box}(\alpha^-\bar{\alpha}^i+\alpha^i\bar{\alpha}^+)\,.
$$
In deriving
of (\ref{intkp}) the value of $\Delta$ given in (\ref{candim}) should be
used.  Finally we are going to choose a basis in which the spin operator
$M^i$ and the generator $K^+$ take the form given in
(\ref{mirep1}),(\ref{mirep2}) and (\ref{cftkp}).  To this end we use the
following decomposition of operator $|{\cal O}\rangle^{iv}$

\begin{equation}\label{oos1os2}
|{\cal O}\rangle^{iv}
=\left(
\begin{array}{l}
|{\cal O}_s^{(1)}\rangle^{iv}
\\[7pt]
|{\cal O}_{s-1}^{(2)}\rangle^{iv}
\end{array}\right)\,,
\end{equation}
where $|{\cal O}_s^{(1)}\rangle^{iv}$ and
$|{\cal O}_{s-1}^{(2)}\rangle^{iv}$
are traceful rank $s$ and $s-1$ tensors of $so(d-3)$ algebra.  In such
basis for $|{\cal O}\rangle^{iv}$ the operator $M^i$
takes the form

\begin{equation}\label{cftapmi}
M^i=\sqrt{\Box}\Bigl((\bar{\alpha}^i+\alpha^i\bar{\alpha}_j^2)\sigma_-
+\alpha^i\sigma_+\Bigr)\,.
\end{equation}
Decomposing $|{\cal O}_{s,s-1}^{(1,2)}\rangle^{iv}$ into
irreducible components we have

$$
|{\cal O}_s^{(1)}\rangle^{iv}
=\sum_{s^\prime}(\alpha_j^2)^{[\frac{s-s^\prime}{2}]}
|{\cal O}_{s^\prime}^{(1)}\rangle^{iv}\,,
\qquad
|{\cal O}_{s-1}^{(2)}\rangle^{iv}
=\sum_{s^\prime}(\alpha_j^2)^{[\frac{s-s^\prime-1}{2}]}
|{\cal O}_{s^\prime}^{(2)}\rangle^{iv}\,,
$$
where $s^\prime$ takes values given in (\ref{phi1decapp}) and
(\ref{phi2decapp}) respectively.
Inserting these expressions into (\ref{oos1os2}) and by applying $M^i$
(\ref{cftapmi}) to (\ref{oos1os2}) we get the following representation of
$M^i$ on $|{\cal O}^{(1,2)}_{s^\prime}\rangle^{iv}$

\begin{eqnarray}
\label{mi1}
{}~&&\hspace{-1.5cm}
 \Box^{-1/2}M^i|{\cal O}_{s^\prime}^{(1)}\rangle^{iv}
=(\alpha^i-\frac{\alpha_j^2\bar{\alpha}^i}{2s^\prime+d-7})
|{\cal O}_{s^\prime-1}^{(2)}\rangle^{iv}
+\frac{1}{2s^\prime+d-3}\bar{\alpha}^i
|{\cal O}_{s^\prime+1}^{(2)}\rangle^{iv}\,,
\\
\label{mi2}
{}~&&\hspace{-1.5cm}
 \Box^{-1/2}M^i|{\cal O}_{s^\prime}^{(2)}\rangle^{iv}
=a(s,s^\prime)(\alpha^i-\frac{\alpha_j^2\bar{\alpha}^i}{2s^\prime+d-7})
|{\cal O}_{s^\prime-1}^{(1)}\rangle^{iv}
+\frac{a(s,s^\prime+1)}{2s^\prime+d-3}
\bar{\alpha}^i|{\cal O}_{s^\prime+1}^{(1)}\rangle^{iv}\,.
\end{eqnarray}
Then the final basis $|{\cal O}_{s^\prime}^{(1,2)}\rangle^{iv}$ in which
the operator $M^i$ and the generator $K^+$ take desired form given in
(\ref{mirep1}), (\ref{mirep2}), (\ref{cftkp}) is found to be

\begin{equation}\label{o4of}
|{\cal O}_{s^\prime}^{(1,2)}\rangle^{iv}
=\sqrt{\Box}^{\,(s-\alpha^i\bar{\alpha}^i)}
|{\cal O}_{s^\prime}^{(1,2)}\rangle\,.
\end{equation}

Now let us discuss the representation of conformal algebra generators
in the space of shadow operators.  To this end it turns out to be
convenient to start with the form of generators given in
$|{\cal O}\rangle^{iv}$ basis (see (\ref{intkp}),(\ref{intjmi})).
First of all taking into account the relationship
(\ref{candim}) the generator $K^+$ (\ref{intkp}) can be rewritten as

\begin{equation}\label{intkpd}
K^+=K_0+\Delta x^+
+\frac{\partial^+}{2\Box} ((\Delta-2)(\Delta+3-d)+\frac{1}{2}M_{ij}^2)\,.
\end{equation}
Then because of relations
$$
(\Delta-2)(\Delta+3-d)
=(\tilde{\Delta}-2)(\tilde{\Delta}+3-d)\,,
\qquad
(K_0^+)^\dagger =-K_0^+ - (d-1) x^+\,,
$$
where $\tilde{\Delta}$ is a conformal dimension of shadow operator
$\tilde{\Delta}=2-s$ we get
$$
(K^+)^\dagger=-\tilde{K}^+\,,
\qquad
\tilde{K}^+\equiv K_0^+ +\tilde{\Delta}x^+
+\frac{\partial^+}{2\Box}
((\tilde{\Delta}-2)(\tilde{\Delta}+3-d)+\frac{1}{2}M_{ij}^2)\,.
$$
From these expressions it is seen that the new generator $\tilde{K}^+$ is
obtainable from $K^+$ (\ref{intkpd}) by making there the substitution
$\Delta\rightarrow \tilde{\Delta}$. This suggests that $\tilde{K}^+$
gives rise representation in the space of shadow operator $|\tilde{\cal
O}\rangle$.  Additional support to this suggestion is that the following
scalar product

$$
\int d^{d-1}x \langle \tilde{\cal O}(x) ||{\cal O}(x)\rangle
$$
is invariant of conformal algebra transformations provided the
$\tilde{{\cal O}}$ is transformed by $\tilde{K}^+$. By introducing new
basis similar to (\ref{oos1os2}), (\ref{o4of}) one finds the representation
of conformal algebra generators in space of shadow operator. These
generators can be obtained from (\ref{cftpa})-(\ref{cftkp}) making there
the substitution $\hat{\Delta}\rightarrow \tilde{\hat{\Delta}}$. Note that
for the case of shadow operator the expression $s-\alpha^i\bar{\alpha}^i$
in r.h.s. of (\ref{o4of}) should be replaced by
$\alpha^i\bar{\alpha}^i-s$\,.

\end{document}